\documentclass[lettersize,journal]{IEEEtran}

\usepackage[T1]{fontenc}
\usepackage{multirow}
\usepackage[table]{xcolor}
\usepackage{cite}
\usepackage{balance}
\usepackage{amsmath,amssymb,amsfonts}
\usepackage{graphicx}
\usepackage{textcomp}
\usepackage{xcolor}
\usepackage{bm}
\usepackage{subfigure}
\usepackage[none]{hyphenat}	
\usepackage{stfloats}
\usepackage{flushend}
\usepackage{algorithm}
\usepackage{algpseudocode}

\DeclareMathOperator*{\argmax}{arg\,max}

\definecolor{electricyellow}{rgb}{1.0, 1.0, 0.0}

\begin{document}
	\bstctlcite{IEEEexample:BSTcontrol}
	
	\title{
		Nature-Inspired Intelligent $\alpha$-Fair
		Hybrid Precoding in Multiuser
		Massive Multiple-Input Multiple-Output Systems
	}
	
	\author{
		Asil Koc,~\IEEEmembership{Graduate Student Member,~IEEE,}
		Tho Le-Ngoc,~\IEEEmembership{Life Fellow,~IEEE}
		\vspace{-3.5ex}
		\thanks{
			This work was partially supported by InterDigital Inc. and the Natural Sciences and Engineering Research Council of Canada (NSERC).
		}
		\thanks{
			A. Koc and T. Le-Ngoc are  with the Department of Electrical and Computer 	Engineering, McGill University, Montréal, QC H3A 0E9, Canada (e-mail:	asil.koc@mail.mcgill.ca; tho.le-ngoc@mcgill.ca). 

			This work has been submitted to the IEEE for possible publication.
			
		}
	}
	
\maketitle	
	
	\begin{abstract}
	This paper proposes a novel nature-inspired $\alpha$-fair hybrid precoding (NI-$\alpha$HP) technique for millimeter-wave multi-user massive multiple-input multiple-output systems.
Unlike the existing HP literature, we propose to apply $\alpha$-fairness for maintaining  various fairness expectations (e.g., sum-rate maximization, proportional fairness, max-min fairness, etc.).
	After developing the analog RF beamformer via slow time-varying angular information, the digital baseband (BB) precoder is designed via the reduced-dimensional effective channel matrix seen from the BB-stage.
For the $\alpha$-fairness, we derive the optimal digital BB precoder expression with a set of parameters, {where optimizing them is an NP-hard problem. Hence, we efficiently optimize the parameters in the digital BB precoder via five nature-inspired intelligent algorithms.}
	Numerical results present that when the sum-rate maximization is the target, the proposed NI-$\alpha$HP technique greatly improves the sum-rate capacity and energy-efficiency performance compared to other benchmarks.
Moreover, NI-$\alpha$HP supports different fairness expectations and reduces the rate gap among UEs by varying the fairness level ($\alpha$).
	\end{abstract}
	

\section{Introduction}

\IEEEPARstart{F}{ifth}-generation (5G) cellular communication networks are rolling out worldwide for supporting ever-growing mobile traffic rates and various user/application demands. 
	Massive multiple-input multiple-output (mMIMO) has already become a key enabling technology in the 5G networks \cite{5G_Mas_MIMO_6,5G_Mas_MIMO_mmWave_5,5G_Book}.
The third generation partnership project (3GPP) standardizes the deployment of up  to 256 antennas at the base station (BS) in Release 17 \cite{Report_5G_Rel17_Mt_256}.	
	Towards the next-generation communication networks, the sixth-generation (6G), the expected impacts of mMIMO systems are further extended with millimeter-wave (mmWave) communications \cite{6G_Docomo}.
In addition to the extremely wide bandwidth at the mmWave frequencies, the shorter wavelengths enable the utilization of larger antenna arrays under the practical area requirement.
	Furthermore, the large antenna arrays can combat the limited scattering mmWave propagation and focus the signal energy in the desired direction via three-dimensional (3D) beamforming \cite{mmWave_Survey}. 
Hence, the mmWave mMIMO technology brings a compelling potential for various emerging applications 
	(e.g., autonomous driving, healthcare, online gaming, augmented/virtual reality (AR/VR), smart home/city, industrial automation, etc.) \cite{6G_prospective_look}.

\subsection{Literature Review}
	Regarding the multi-user (MU) downlink transmission, the precoding is an essential signal processing technique for reliable communication quality.
In the conventional MIMO systems, the single-stage fully-digital precoding (FDP) is widely studied thanks to the limited number of antennas \cite{MassMIMO_Precoding_2021}.
	Nevertheless, FDP poses two critical issues for the mMIMO systems: (i) high power consumption as well as hardware cost/complexity due to a dedicated power-hungry RF chain per antenna, (ii) large channel estimation overhead size.
Hybrid precoding (HP) is proposed as a promising technique, which splits the precoder into two stages as RF-stage and baseband(BB)-stage
\cite{Mass_MIMO_Hybrid_Survey,Mass_MIMO_Hyb_Survey,Mass_MIMO_Hybrid_Survey_2}.
	In comparison to the large antenna array size, the HP architecture employs a significantly low number of RF chains \cite{MassMIMO_hybrid_NO_ADMA}.
HP is also capable of reducing the channel estimation overhead size, when the RF-stage is developed via slow time-varying channel state information (CSI) (e.g., angle-of-departure (AoD), channel covariance matrix) \cite{JSDM_LargeArray,ASIL_EE_2D_OJ_COMS,ASIL_ABHP_Access,ASIL_FD_MU_NOBF}.
	For instance,  eigen-beamforming based HP (EBF-HP) proposed in \cite{JSDM_LargeArray} uses the channel covariance matrix in the RF-stage design.
EBF-HP is also investigated for various antenna array geometries and low-resolution hardware components in \cite{ASIL_EE_2D_OJ_COMS}.
	Afterwards, angular-based HP (AB-HP) is introduced to design the RF-stage via AoD parameters in \cite{ASIL_ABHP_Access}. The authors present that AB-HP achieves a higher sum-rate capacity than EBF-HP
		while closely approaching the sum-rate capacity of the conventional FDP technique.
Then, AB-HP is investigated for full-duplex communications in \cite{ASIL_FD_MU_NOBF}, where the BS operates in full-duplex for the simultaneous downlink and uplink transmission.

Fairness is principally an abstract sociopolitical concept implying justice, equity, and impartiality in the distribution of finite resources among users/clients in a given system \cite{Fairness_Survey_1}.
When the technical systems are studied, such as wireless communications, the fairness is attributed to the fair allocation of the valuable resources (e.g., power and bandwidth)
\cite{Fairness_Survey_2}.
\begin{table*}
	\caption{State-of-the-Art for Precoding Techniques in Massive MIMO Systems.}
	\label{table_LiteratureReview}
	\footnotesize 
	\centering
	\renewcommand{\arraystretch}{1.2}
	\begin{tabular}{|c||c|c|c|c|c|c|c|}
		\hline
		\multirow{2}{*}{\hspace{-1ex}Reference\hspace{-1ex}} & 
		Multi&
		Hybrid&
		Low CSI&
		Max-Min&
		Proportional&
		Sum-Rate&
		\multirow{2}{*}{\hspace{-1ex}$\alpha$-Fairness\hspace{-1ex}} 
		\\
		&
		User&
		Precoding&
		Overhead&
		Fairness&
		Fairness&
		Maximization&
		\\
		\hline
		\hline
		\cite{MassMIMO_hybrid_NO_ADMA}&
		\checkmark&
		\checkmark&&&&&
		\\
		\hline
		\cite{JSDM_LargeArray,ASIL_EE_2D_OJ_COMS,ASIL_ABHP_Access,ASIL_FD_MU_NOBF} &
		\checkmark&
		\checkmark&
		\checkmark&&&&
		\\
		\hline
		\cite{MaxMinFairness,MMF_NOMA_Melda_2021}&
		\checkmark&&&
		\checkmark	&&	&
		\\
		\hline
		\cite{MaxMin_HP_Maritime}&
		\checkmark&
		\checkmark&&
		\checkmark&&&
		\\
		\hline
		\cite{proportional_Fairness}&
		\checkmark&
		&
		&
		&
		\checkmark&&
		\\
		\hline
		\cite{HBF_fairness_proportional}&
		\checkmark&
		\checkmark&
		&
		&
		\checkmark&&
		\\
		\hline
		\cite{SumRateMax_NOMA_LEO}&
		\checkmark&&&&&
		\checkmark&
		\\
		\hline
		\cite{ASIL_PSO_PA_WCNC,ASIL_DL_PA_ICC,ASIL_GA_RA_OFDM_WCNC} &
		\checkmark&
		\checkmark&
		\checkmark&&&
		\checkmark&
		\\
		\hline
		\cellcolor{yellow} \hspace{-1ex}This Paper\hspace{-1ex}
		&\cellcolor{yellow}\checkmark
		&\cellcolor{yellow}\checkmark
		&\cellcolor{yellow}\checkmark
		&\cellcolor{yellow}\checkmark
		&\cellcolor{yellow}\checkmark
		&\cellcolor{yellow}\checkmark
		&\cellcolor{yellow}\checkmark
		\\ \hline
	\end{tabular}
\end{table*}
	There are various qualitative fairness measures, such as  
	max-min fairness \cite{MaxMinFairness,MMF_NOMA_Melda_2021,MaxMin_HP_Maritime},
	proportional fairness \cite{proportional_Fairness,HBF_fairness_proportional},
	and
	sum-rate maximization \cite{SumRateMax_NOMA_LEO,ASIL_GA_RA_OFDM_WCNC,ASIL_PSO_PA_WCNC,ASIL_DL_PA_ICC}.
Particularly, \cite{MaxMin_HP_Maritime} targets the maximization of minimum user rate during the HP design assuming the availability of full-size CSI. 
	However, in general, the max-min fairness might lead to sum-rate capacity degradation in the case of a weak user.
On the contrary, \cite{SumRateMax_NOMA_LEO,ASIL_GA_RA_OFDM_WCNC,ASIL_PSO_PA_WCNC,ASIL_DL_PA_ICC} aims to maximize the sum-rate capacity in the MU-mMIMO systems, where FDP and HP are studied in \cite{SumRateMax_NOMA_LEO} and \cite{ASIL_GA_RA_OFDM_WCNC,ASIL_PSO_PA_WCNC,ASIL_DL_PA_ICC}, respectively.
	Nonetheless, the sum-rate maximization objective might result in poor performance for a few users, while enhancing the overall system sum-rate capacity.
To exploit the best of two worlds, $\alpha$-fairness is an impressive concept to satisfy various fairness expectations by adjusting a single non-negative scalar, denoted as $\alpha\ge 0$ \cite{alphaFAIR_1_NOMA}.
	The larger value of $\alpha$ signifies increased fairness among the users. For instance, $\alpha=0$ indicates the sum-rate maximization objective, whereas the proportional and max-min fairness are represented by $\alpha=1$ and $\alpha\to\infty$, respectively.

	As a powerful component of artificial intelligence (AI), nature-inspired optimization algorithms have recently gained considerable attention by means of their success \cite{PSO_AI}.	
Unlike the traditional optimization techniques, the nature-inspired algorithms enjoy the utilization of intelligent search agents, which follow stochastic and deterministic components \cite{PSO_book}.
	As it is inspired by the nature, there are several ways to implement the characteristics of search agents, such as 
	particle swarm optimization (PSO) \cite{PSO_AI},
	grey wolf optimization (GWO) \cite{GWO_first_paper},
	ant colony optimization (ACO) \cite{ACO_cont},
	cuckoo search (CS) \cite{PSO_book},
	firefly algorithm (FA) \cite{PSO_book}, etc.
As stated by the \textit{no-free-lunch theorem},  there is no universally efficient algorithm valid for all types of optimization problems \cite{NFL_first_TEVC}.
	In other words, the performance of each nature-inspired algorithm heavily depends on the problem itself.
For example, a comprehensive comparison of various nature-inspired intelligent algorithms (e.g., including GWO, PSO, CS) is presented in \cite{UAV_swarm_intelligence}, where GWO achieves the best performance on the location optimization of a drone BS.
	On the other hand, a PSO-based power allocation (PSO-PA) algorithm along with AB-HP is proposed for the MU-mMIMO systems in \cite{ASIL_PSO_PA_WCNC}. 
	It is demonstrated that PSO-PA tightly approaches the optimal PA scheme, when it is compared with the computationally complex exhaustive search.
Then, PSO-PA is used to develop a supervised deep learning mechanism in \cite{ASIL_DL_PA_ICC}, which makes both robust and faster predictions for the allocated powers.
	
\subsection{Contributions}
A new nature-inspired intelligent $\alpha$-fair HP (NI-$\alpha$HP) technique is proposed for the Wave MU-mMIMO systems in this paper.
	Different from the existing HP techniques in the literature, $\alpha$-fairness objective is employed during the HP design for achieving various fairness requirements.
Table \ref{table_LiteratureReview} summarizes a detailed comparison of this paper and existing precoding techniques in the mMIMO systems.

The main contributions of this paper are summarized as:
\begin{itemize}
	\item \textbf{Nature-Inspired $\alpha$-Fair Hybrid Precoding (NI-$\alpha$HP):}
	We introduce a two-stage NI-$\alpha$HP technique, where
	our main goals include
		performing $\alpha$-fair resource allocation,
	reducing the number of RF chains, and lowering the CSI overhead size.
		Analog RF beamformer, as the first-stage, is developed via slow time-varying AoD information to reduce the CSI overhead size and maximize the beamforming gain in the desired direction.
	Digital BB precoder, as the second-stage, is constructed via the reduced-dimensional effective channel matrix. 
	According to the $\alpha$-fairness objective, we obtain an optimal digital BB precoder expression including a set of {NP-hard} parameters.
	Five nature-inspired intelligent algorithms (i.e., PSO, GWO, ACO, CS, FA) are employed to optimize these parameters.
	{Here, we aim to investigate their performance and efficiency in our problem.}
	\item \textbf{Low CSI Overhead \& Hardware Cost/Complexity:} 
	The instantaneous channel estimation overhead size is remarkably lowered by using the slow time-varying AoD information at the RF-stage.
		Furthermore, the analog RF beamformer and digital BB precoder are interconnected via a significantly low number of RF chains compared to the large antenna array size.
	The numerical results show that when the BS is equipped with $256$ antennas, the proposed NI-$\alpha$HP technique is capable of reducing the number of RF chains and CSI overhead size by $93.75\%$.
	\item \textbf{Sum-Rate \& Energy-Efficiency:}
Regarding the sum-rate maximization objective, we present that the proposed NI-$\alpha$HP technique greatly enhances the sum-rate capacity compared to other HP benchmarks. Furthermore, NI-$\alpha$HP achieves considerably higher energy-efficiency than the conventional FDP by means of reduced hardware cost/complexity with a low number of RF chains.
	Also, we observe that GWO converges faster and achieves higher capacity among all five NI-$\alpha$HP techniques.
	\item \textbf{Fairness:}
	Although the sum-rate maximization leads to a high rate gap among the users, 
	NI-$\alpha$HP successfully decreases the rate gap and satisfies various fairness expectations by only adjusting a single scalar (i.e., fairness level $\alpha$).
		Also, Jain's fairness index is employed as a quantitative measure. 
	We show that as the fairness level $\alpha$ increases,  NI-$\alpha$HP improves Jain's fairness index.
\end{itemize}

\subsection{Organization}
The rest of this paper is organized as follows. 
	Section \ref{sec_SYS} introduces the system model for the mmWave MU-mMIMO systems.
Section \ref{sec_ProblemForm} expresses the problem formulation on $\alpha$-fairness.
	Afterwards, we propose five NI-$\alpha$HP techniques in Section \ref{sec_HP}.
Then, Section \ref{sec_Results} presents the comprehensive illustrative results.
	Finally, Section \ref{sec_Conc} concludes this paper.

\subsection{Notation}
Bold upper/lower case letters denote matrices/vectors. 
	$\left( \cdot \right)^*$, 
	$\left( \cdot \right)^T$, 
	$\left( \cdot \right)^H$,
	$\left\| \cdot \right\|$, 
		and 
	$\left\| \cdot \right\|_F$ represent the complex conjugate, the transpose, the conjugate transpose, the $2$-norm, and the Frobenius norm, respectively. 
$\mathbf{I}_K$, ${\mathbb{E}}\left\lbrace \cdot\right\rbrace $, $\rm{tr}\left(\cdot\right)$, and $\angle \left(\cdot\right)$ stand for $K\times K$ identity matrix, the expectation operator, the trace operator, and the argument of a complex number, respectively. 
${\bf X}[m,n]$	denotes the element at the intersection of $m^{th}$ row  and $n^{th}$ column.
$\mathbf{X}\otimes\mathbf{Y}$ and $\mathbf{X}\odot\mathbf{Y}$ are respectively the Kronecker and Hadamard product of two matrices $\mathbf{X}$ and $\mathbf{Y}$.
We use ${{x}}\sim{\cal C}{\cal N}\left( {0 ,{\sigma}} \right)$, when ${x}$ is a complex Gaussian random variable with zero-mean and variance ${\sigma}$.
Also, $x\sim\textrm{Ber}\left(\sigma\right)$ represents a Bernoulli random variable, which equals $x=1$ with the probability of $\sigma\in\left[0,1\right]$, otherwise, $x=0$.

\section{System Model}\label{sec_SYS}
\begin{figure}[!t]
	\centering
	\includegraphics[width=\columnwidth]
	{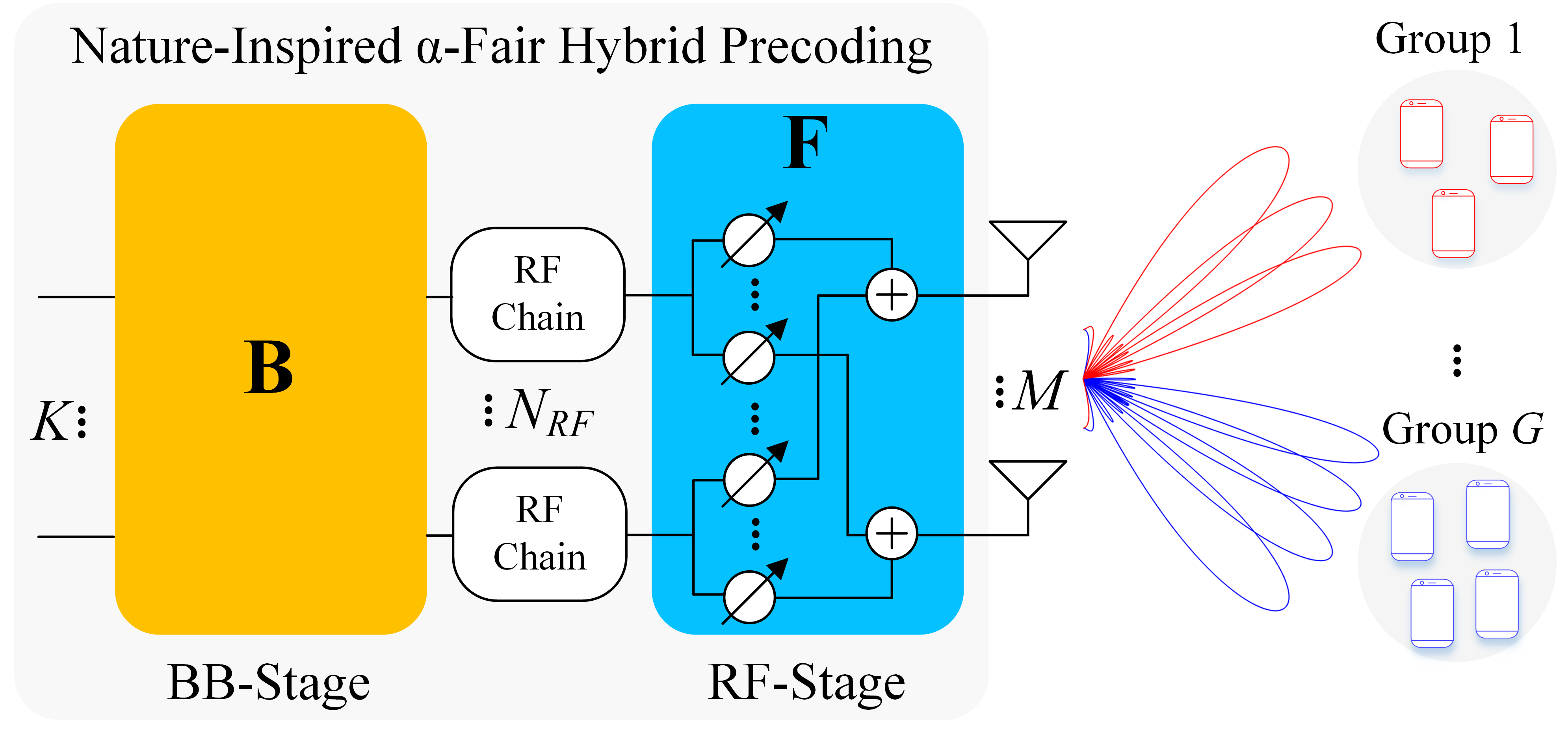}
	\caption{MU-mMIMO systems with $\alpha$-fair hybrid precoding ($\alpha$HP).}
	\label{fig_1_SystemModel}
\end{figure}

A single-cell MU-mMIMO system is considered for the downlink transmission as illustrated in Fig. \ref{fig_1_SystemModel}, where a base station (BS) with $M$ antennas serves $K$ single-antenna user equipments (UEs) clustered in $G$ groups. 
	The BS employs a uniform rectangular array (URA) with $M=M_x\times M_y$ antennas to enable 3D beamforming by exploiting azimuth and elevation angles, where $M_x$ and $M_y$ denotes the number of antennas along $x$-axis and $y$-axis, respectively \cite{5G_Mas_MIMO_mmWave_5}. 
Unlike the widely considered uniform linear array (ULA), URA packs a large number of antennas on a two-dimensional (2D) grid and enhances the area efficiency for the practical deployment of mMIMO systems  
\cite{ASIL_EE_2D_OJ_COMS,5G_Mas_MIMO_mmWave_5,  Mass_MIMO_Hyb_Survey}.
	Furthermore, we assume that there are $K_g$ UEs in the $g^{th}$ group with $K=\sum_{g=1}^{G}K_g$.

According to the HP architecture illustrated in Fig. \ref{fig_1_SystemModel}, the transmitted downlink signal is defined as 
	${{\bf s}={\bf FBd}\in{\mathbb{C}^M}}$, where
	${{\bf F}\in\mathbb{C}^{M\times N_{RF}}}$
is the analog RF beamformer matrix,
	${{\bf B}=\left[{\bf b}_1,\cdots,{\bf b}_K\right]\in\mathbb{C}^{N_{RF} \times K}}$
is the digital BB precoder matrix,
	${\bf d}=\left[d_1,\cdots,d_K\right]\in\mathbb{C}^{K}$
is the data signal vector encoded by i.i.d. Gaussian codebook with the entries following the distribution of $\mathcal{CN}\left(0,1\right)$ (i.e., $\mathbb{E}\big\{{\bf d d}^H\big\}={\bf I}_K$).
	Here, $N_{RF}$ denotes the number of RF chains for interconnecting the RF-stage and BB-stage.
It is important to highlight that $N_{RF}$ is chosen based on the following condition $K\le N_{RF} \ll M$ to support $K$ single-antenna UEs, while remarkably reducing the hardware cost/complexity and power consumption in the MU-mMIMO system with large antenna arrays.
	Additionally, the analog RF beamformer is constructed via low-cost phase-shifters, which brings the constant modulus (CM) constraint (i.e., $\big|{\bf F}[{i,j}]\big|=\frac{1}{\sqrt{M}}, \forall i,j,$)
	Furthermore, the transmitted signal satisfies the maximum transmit power constraint of $P_T$ (i.e., $\mathbb{E}\big\{\big\|{\bf s}\big\|^2\big\}\le P_T$).

The instantaneous downlink channel matrix is defined as 
	${\bf H}=\left[{\bf h}_1,\cdots,{\bf h}_K\right]^T\in\mathbb{C}^{K\times M}$ 
with 
	${\bf h}_k\in\mathbb{C}^{M}$ as the $k^{th}$ UE channel vector.
	Different from the sub-6 GHz frequency bands, the mmWave channels experience a limited scattering propagation environment	\cite{mmWave_Survey}.
By employing the 3D geometry-based stochastic channel model for mmWave communications \cite{ChannelModels,COST_2100_MIMO,GBSM_3GPP}, the $k^{th}$ UE channel is given by:
\begin{equation}\label{eq_h_k}
	{\bf{h}}_k
	=
	\sum_{l=1}^{L} {\tau^{-\eta}_{k_l}} {z_{k_l}} {\bm{\phi }}
	\big(
	{{\gamma _{x,k_l}},{\gamma _{y,k_l}}}
	\big) 
	= {{\bf{\Phi }}}_k^T{\bf{z}}_k
	\in
	\mathbb{C}^M,
\end{equation}
where 
$L$ is the number of paths,
$\eta$ is the path loss exponent,
$\tau_{k_l}$ and $z_{k_l}\sim\mathcal{CN}\big(0,{1}/{L}\big)$ are the distance and complex path gain of $l^{th}$ path, respectively,
${\bm{\phi }}\big( \cdot,\cdot\big)\in \mathbb{C}^M$ is the phase response vector,
$\gamma_{x,k_l}=\sin\left(\theta_{k_l}\right)\cos\left(\psi_{k_l}\right)\in\left[-1,1\right]$
and
$\gamma_{y,k_l}=\sin\left(\theta_{k_l}\right)\sin\left(\psi_{k_l}\right)\in\left[-1,1\right]$
are the coefficients reflecting the elevation AoD (EAoD) and azimuth AoD (AAoD) for the corresponding path.
Here,
we define
$\theta_{k_l}\in\big[\theta_k - \delta_k^\theta, \theta_k + \delta_k^\theta\big]$ as the EAoD with mean $\theta_k$ and spread $\delta_k^\theta$, 
$\psi_{k_l}\in\big[\psi_k - \delta_k^\psi, \psi_k + \delta_k^\psi\big]$ as the AAoD with mean $\psi_k$ and spread $\delta_k^\psi$.
Based on the URA structure, we define the phase response vector as \cite{ASIL_ABHP_Access}:
\begin{equation}\label{eq_phase_vector}
	\begin{aligned}
		{\bm{\phi}}\hspace{-0.5ex}\left( {{\gamma_x, \gamma_y}} \right) \hspace{-0.5ex}&=\hspace{-0.75ex} \big[ {1,{e^{ -j2\pi d  {{\gamma_x }} }}, \cdots,{e^{ -j2\pi d\left( {{M_{x}} - 1} \right) {{\gamma_x}} }}} \big]^T\\
		&\otimes \hspace{-0.5ex}\big[ {1,{e^{-j2\pi d  {{\gamma_y }} }}, \cdots,{e^{-j2\pi d\left( {{M_{y}} - 1} \right) {{\gamma_y}} }}} \big]^T\in \mathbb{C}^M,
	\end{aligned}
\end{equation}
where $d=0.5$ is the normalized half-wavelength distance between neighboring antenna elements.
As shown in \eqref{eq_h_k}, the instantaneous channel vector is composed of two main components: (i) the fast time-varying path gain vector 
${{\bf z}_k = \big[\tau^{-\eta}_{k_1}z_{k_1},\cdots,\tau^{-\eta}_{k_L}z_{k_K}\big]^T\in\mathbb{C}^L}$,
(ii) slow time-varying phase response matrix 
${{\bf{\Phi }}}_k\in\mathbb{C}^{L\times M}$ as a function of AoD information.

The received signal at the $k^{th}$ UE is written as follows:
\begin{equation}
	\begin{aligned}
		r_k
		&
		=
		{\bf h}_k^T {\bf s} + n_k
		\\
		&
		=
		\underbrace{{\bf h}_k^T {\bf Fb}_kd_k}_{\textrm{Intended Signal}}
		+
		\underbrace{\sum\nolimits_{u\ne k}^{K}{\bf h}_k^T {\bf Fb}_u d_u}_{\textrm{Inter UE Interference}}
		+ 
		\underbrace{n_k}_{\textrm{Noise}},
	\end{aligned}
\end{equation}
where
$n_{k}\sim\mathcal{CN}\big(0,\sigma_n^2\big)$ is the  circularly symmetric complex Gaussian noise. After some mathematical manipulations, the instantaneous signal-to-interference-plus-noise-ratio (SINR) at the $k^{th}$ UE is obtained as:
\begin{equation}\label{eq_SINR}
	\begin{aligned}
		\textrm{SINR}_{k}
		\hspace{-0.35ex}
		\left(
		{\bf F},
		{\bf B}
		\right)
		=
		\frac
		{\left|{\bf h}_k^T{\bf Fb}_k\right|^2}
		{\sum_{u\ne k}^{K}\left|{\bf h}_k^T{\bf Fb}_u\right|^2 
			+
		\sigma_n^2}.
	\end{aligned}
\end{equation}
Afterwards, as a function of the analog RF beamformer ${\bf F}$ and digital BB precoder ${\bf B}$, the $k^{th}$ UE rate is calculated as:
\begin{equation}\label{eq_Rk}
	R_k
	\hspace{-0.5ex}
	\left({\bf F},{\bf B}\right)
		\hspace{-0.25ex}=\hspace{-0.25ex}
	\log_2
	\hspace{-0.5ex}
	\left(
	\hspace{-0.25ex}
	1
		\hspace{-0.25ex}+\hspace{-0.25ex}
	\frac{\left|{\bf h}_k^T{\bf Fb}_k\right|^2}{\sum_{u\ne k}^{K}\hspace{-0.25ex}\left|{\bf h}_k^T{\bf Fb}_u\right|^2 	
		\hspace{-0.25ex}+\hspace{-0.25ex}
	\sigma_n^2}
	\hspace{-0.25ex}
	\right)
	\hspace{-0.5ex}
	\textrm{ [bps/Hz]}.
\end{equation}
Finally, we find the sum-rate capacity in the MU-mMIMO systems as 
	$R_{\textrm{sum}}\left({\bf F},{\bf B}\right) = \sum_{k=1}^{K}R_k\left({\bf F},{\bf B}\right)$.

\section{Problem Formulation on $\alpha$-Fairness}\label{sec_ProblemForm}

According to \eqref{eq_SINR} and \eqref{eq_Rk}, we aim to jointly maximize the intended signal power 
$\left|{\bf h}_k^T{\bf Fb}_k\right|^2$, $\forall k$,
and
mitigate the inter UE interference power
$\left|{\bf h}_k^T{\bf Fb}_u\right|^2$, $\forall u\neq k$ in the HP design.
Under the maximum transmit power constraint of $P_T$, we have 
$\mathbb{E}\big\{\big\|{\bf s}\big\|^2\big\}
=
\sum_{k=1}^{K} {\bf b}_k^H {\bf F}^H {\bf F} {\bf b}_k
=
\sum_{k=1}^{K} p_k
\le P_T$,
where 
$p_k = {\bf b}_k^H {\bf F}^H {\bf F} {\bf b}_k$ is the allocated power for the $k^{th}$ UE during the HP design.
	However, while allocating the limited resources (e.g., power) among the downlink UEs, there is an interesting trade-off between the sum-rate maximization and fairness.
For example, by defining $R_{\max} =\max_k R_k$ and $R_{\min} = \min_k R_k$, the sum-rate maximization objective only focuses on optimizing $R_{\textrm{sum}}= \sum_{k=1}^{K}R_k$, which might cause unfair UE rates with a large gap between $R_{\max}$ and $R_{\min}$.
	Moreover, it might contradict the quality of service (QoS) requirement (e.g., achieving a minimum rate for each UE).
On the other hand, as an extreme opposite scenario, the max-min fairness only targets maximizing $R_{\min} = \min_k R_k$, which can enhance the QoS requirements. Nonetheless, it might lead to a considerable degradation in the overall system capacity with a greatly reduced sum-rate $R_{\textrm{sum}}$ \cite{MaxMinFairness}.

For this purpose, 
we here propose to apply $\alpha$-fairness, which utilizes a single scalar for enabling different UE fairness levels. In order to model $\alpha$-fairness in the MU-mMIMO systems, we first define the following utility function\cite{Fairness_Survey_2}:
\begin{equation}\label{eq_alpha_fair}
	\mathcal{U}_{\alpha}\left(x\right)=
	\left\{ {\begin{array}{*{20}{l}}
			\frac{x^{1-\alpha}}{1-\alpha},&{\alpha\ge 0, \alpha \ne 1}\\
			\ln\left(x\right),& \alpha = 1,
	\end{array}} \right.
\end{equation}
where $\alpha$ represents different fairness levels and $x\ge 0$ could be considered as a given UE rate. By using \eqref{eq_Rk} and \eqref{eq_alpha_fair}, we formulate the objective function on $\alpha$-fair HP ($\alpha$HP) design in the MU-mMIMO systems as follows:
\begin{equation}\label{eq_obj_1}
	\begin{aligned}
		\max_{{\bf F},{\bf B}}
		&
		\sum_{k=1}^{K}
		\mathcal{U}_{\alpha}\left(R_k\right),\\
		\textrm{s.t. }
		&
		C_1:~ 
		R_k =\log_2\left( 1 + \textrm{SINR}_k \right),\\
		&
		C_2:~ 
		\textrm{SINR}_{k}
		\hspace{-0.35ex}
		=
		\frac
		{\left|{\bf h}_k^T{\bf Fb}_k\right|^2}
		{\sum_{u\ne k}^{K}\left|{\bf h}_k^T{\bf Fb}_u\right|^2 
			+
		\sigma_n^2},\\
		&
		C_3:~ 
		\mathbb{E}\big\{\big\|{\bf s}\big\|_2^2\big\}=\sum_{k=1}^{K}{\bf b}_k^H{\bf F}^H {\bf F} {\bf b}_k=\sum_{k=1}^{K}p_k\le P_T,\\
		&
		C_4:~
		p_k\ge 0, \forall k,
		\\
		&
		C_5:~ 
		\big|{\bf F}[{i,j}]\big|=\frac{1}{\sqrt{M}}, \forall i,j,
	\end{aligned}
\end{equation}
where $C_1$ and $C_2$ refer the rate and SINR at the $k^{th}$ UE, respectively, $C_3$ implies the total transmit power constraint, $C_4$ indicates the non-negative allocated power for each downlink UE,
and $C_5$ expresses the CM constraint at the analog RF beamformer design due to the utilization of low-cost phase-shifters.
	By adjusting the value of  $\alpha$, the above objective function addresses various fairness levels.
	To illustrate, one can show that:
	(i) 	$\alpha=0$ is for the sum-rate maximization with $\max \sum_{k=1}^{K}R_k$,
	(ii)	$\alpha=1$ is for the proportional fairness with $\max\sum_{k=1}^{K}\ln\left(R_k\right)$,
	(iii)	$\alpha\to\infty$ is for the max-min fairness with $\max\left[\lim\limits_{\alpha\to\infty}\sum_{k=1}^{K}\frac{R_k^{1-\alpha}}{1-\alpha}\right]
		=
	\max \left[\min_k R_k\right]$,
	(iv)	other values of $\alpha$ are for different fairness levels.

However, the objective function given in \eqref{eq_obj_1} is a non-convex optimization problem because of several reasons (e.g., the CM constraint \cite{Mass_MIMO_Hyb_Survey}, the optimization variables interchangeably located in the numerator/denominator \cite{Emil_PA_NonConvex,PA_nonConvex,PA_nonConvex_2}).
	For this reason, in Section \ref{sec_HP}, the proposed HP solution is developed by applying various nature-inspired intelligent algorithms.

\section{Nature-Inspired Intelligent $\alpha$-Fair \\Hybrid Precoding (NI-$\alpha$HP)}\label{sec_HP}

In this section, we introduce the proposed nature-inspired intelligent $\alpha$-fair hybrid precoding (NI-$\alpha$HP) technique for the MU-mMIMO systems. Along with the aim of  optimal $\alpha$-fair resource allocation, our main objectives also include
	reducing the RF chain utilization
		and
	decreasing the channel estimation overhead size.
Hence, the proposed NI-$\alpha$HP technique sequentially develops the analog RF beamformer ${{\bf F}\in\mathbb{C}^{M\times N_{RF}}}$ and digital BB precoder ${{\bf B}\in\mathbb{C}^{N_{RF}\times K}}$. First, we only employ the slow time-varying AoD information\footnote{In addition to the conventional AoD estimation techniques \cite{AoD_Est_2_Decades}, the AoD parameters (e.g., mean and spread) can be acquired via a deep learning and geospatial data-based offline estimation technique presented in \cite{ASIL_Xiaoyi_DL_CE}.} in the analog RF beamformer design to maximize the beamforming gain in the intended direction, while lowering the instantaneous CSI requirements for the MU-mMIMO systems. Second, we derive the optimal parameterized digital BB precoder solution by using the reduced-size effective CSI, where five nature-inspired optimization algorithms are applied to optimize those parameters.

\subsection{Analog RF Beamformer}\label{secsub_RF}

When UEs are clustered in multiple geographical locations as illustrated in Fig. \ref{fig_1_SystemModel}, each UE group experiences similar AoD information \cite{ASIL_Xiaoyi_DL_CE}. Hence, we design the analog RF beamformer matrix with $G$ blocks as follows:
\begin{equation}\label{eq_F}
	{\bf F}= \left[{\bf F}_1,\cdots,{\bf F}_G\right]\in \mathbb{C}^{M\times N_{RF}},
\end{equation}
where 
${\bf F}_g \in \mathbb{C} ^{M\times N_{RF,g}}$
and
$N_{RF,g}$
are the RF beamformer
and 
number of RF chains for the $g^{th}$ UE group, respectively, 
with $N_{RF}=\sum\nolimits_{g=1}^{G}N_{RF,g}$.
	By using \eqref{eq_phase_vector}, the analog RF beamformer is constructed via the unit-power steering vectors defined as follows:
\begin{equation}\label{eq_steering_vec}
	{\bf e}\left(\gamma_x,\gamma_y\right)=\frac{1}{\sqrt{M}}{\bm{\phi}}^*\left( {{\gamma_x, \gamma_y}} \right)\in\mathbb{C}^M, \forall\gamma_x,\gamma_y\in\left[-1,1\right],
\end{equation}
which satisfies ${\left\| {\bf e}\left(\gamma_x,\gamma_y\right) \right\|^2=1}$
and the CM constraint (i.e., $C_5$ in \eqref{eq_obj_1}).
Afterwards, we define the quantized angle-pairs as 
${{\lambda^x_{m}} = -1 + \frac{2m-1}{{{M_{x}}}}}$ 
for 
$m = 1, \cdots,{M_{x}}$
and 
${{\lambda^y_{n}} = -1 + \frac{2n-1}{{{M_{y}}}}}$ 
for
$n = 1, \cdots,{M_{y}}$. Here, we have $M$ possible $\left({\lambda^x_{m}},{\lambda^y_{n}}\right)$ angle-pairs, which enable us to cover the complete 3D elevation and azimuth angular domain with the minimum number of steering vectors \cite{ASIL_ABHP_Access}.
	Moreover, the quantized angle-pairs satisfy the orthogonality property (i.e., 
$
	{\bf e}^H\big(\lambda^{x}_m,\lambda^{y}_n\big){\bf e}\big(\lambda^x_{m'},\lambda^y_ {n'}\big)=0,  
	\forall \left[m,n\right]
	\ne \left[m',n'\right]
$).

	Based on the similarity of AoD information within each UE group, one can define the AoD support of the $g^{th}$ group as:
\begin{equation}\label{eq_AoD_Supp}
	\mathcal{A}_g = 
	\left[
	\begin{array}{*{20}{c}}
		\gamma_x = \sin\left(\theta\right) \cos\left(\psi\right)
		\\
		\gamma_y = \sin\left(\theta\right) \sin\left(\psi\right)
	\end{array}
	\right]
	\Big|
	\forall\theta \in {\bm{\theta}}_g,
	\forall\psi \in {\bm{\psi}}_g,
	\Big.
\end{equation}
where 
$\bm{\theta}_g=\left[\theta_g-\delta_g^\theta, \theta_g+\delta_g^\theta\right]$
and
$\bm{\psi}_g=\left[\psi_g-\delta_g^\psi, \psi_g+\delta_g^\psi\right]$
are the boundaries EAoD and AAoD, respectively. 
Here, $\theta_g$ and $\psi_g$  represent the mean EAoD and AAoD for the corresponding UE group with their spread of $\delta_g^\theta$ and $\delta_g^\psi$, respectively.
	By using \eqref{eq_h_k}, \eqref{eq_phase_vector} and \eqref{eq_AoD_Supp}, the $g^{th}$ UE group channel matrix can be rewritten as follows \cite{JSDM_LargeArray,ASIL_EE_2D_OJ_COMS,ASIL_ABHP_Access,ASIL_FD_MU_NOBF}:
\begin{equation}\label{eq_H_g}
	{\bf H}_g 
	= 
	\big[
	{\bf h}_{g_1},\cdots,{\bf h}_{g_{K_g}}
	\big]^T
	= 
	{\bf Z}_g{\bf\Phi}_g
	\in
	\mathbb{C}^{K_g\times M},
\end{equation}
where 
${\bf Z}_g = \big[
{\bf z}_{g_1},\cdots,{\bf z}_{g_{K_g}}
\big]^T\in\mathbb{C}^{K_g\times L}$ 
is the fast time-varying path gain matrix,
${\bf \Phi}_g\in\mathbb{C}^{L\times M}$ 
is the slow time-varying phase response matrix,
$g_k=k+\sum_{t=1}^{g-1}K_t$ is the UE index.
By using \eqref{eq_F}, \eqref{eq_H_g}, and the concatenated channel matrix as
${\bf H} =
\left[
{\bf H}_1^T,
\cdots,
{\bf H}_G^T
\right]^T
\in\mathbb{C}^{K\times M}$, the effective channel matrix seen from the BB-stage is obtained as:
\begin{equation}\label{eq_H_eff}
	\bm{\mathcal{H}}
	\hspace{-.25ex}=\hspace{-.25ex}
	{\bf H}
	{\bf F} 
	\hspace{-.25ex}=\hspace{-.5ex}
	\left[ \hspace{-1ex}{\begin{array}{*{20}{c}}
			{{{\bf{H}}_1}{{\bf{F}}_1}}&\hspace{-1ex}{{{\bf{H}}_1}{{\bf{F}}_2}}&\hspace{-1ex}{\cdots}&\hspace{-1ex}{{{\bf{H}}_1}{{\bf{F}}_G}}\\
			{{{\bf{H}}_2}{{\bf{F}}_1}}&\hspace{-1ex}{{{\bf{H}}_2}{{\bf{F}}_2}}&\hspace{-1ex}{\cdots}&\hspace{-1ex}{{{\bf{H}}_2}{{\bf{F}}_G}}\\
			{\vdots}&\hspace{-1ex}{\vdots}&\hspace{-1ex}{\ddots}&\hspace{-1ex}{\vdots}\\
			{{{\bf{H}}_G}{{\bf{F}}_1}}&\hspace{-1ex}{{{\bf{H}}_G}{{\bf{F}}_2}}&\hspace{-1ex}{\cdots}&\hspace{-1ex}{{{\bf{H}}_G}{{\bf{F}}_G}}\\
	\end{array}} \hspace{-1ex} \right]\hspace{-.75ex}\in\mathbb{C}^{K\times N_{RF}},
\end{equation}
where $\bm{\mathcal{H}}_g={\bf H}_g{\bf F}_g={\bf Z}_g{\bf \Phi}_g{\bf F}_g\in\mathbb{C}^{K_g\times N_{RF,g}}$ is the effective intended channel matrix for the $g^{th}$ UE group and ${\bf H}_t{\bf F}_g\hspace{-0.5ex}=\hspace{-0.5ex}{\bf Z}_t{\bf \Phi}_t{\bf F}_g\in\mathbb{C}^{K_t\times N_{RF,g}}$ is the effective interference channel matrix among the $g^{th}$ and $t^{th}$ UE group, $\forall t\neq g$.
Hence, we have two main goals for the RF beamformer design: 
(i) maximize the beamforming gain towards the intended direction via constructing the columns of ${\bf F}_g$  from the subspace spanned by ${\bf \Phi}_g$ 
(i.e., $\textrm{Span}\left({\bf F}_g\right) \subset  \textrm{Span}\left({\bf \Phi}_g\right)$),
(ii) suppress the inter-group interference via selecting the columns of ${\bf F}_g$ orthogonal to ${\bf \Phi}_t$, $\forall t\ne g$ 
(i.e., $\textrm{Span}\left({\bf F}_g\right) \subset  \cup_{t\ne g}\textrm{Null}\left({\bf \Phi}_t\right)$).
In other words, the second goal targets the approximate zero condition as:
\begin{equation}\label{eq_approx_zero}
	{\bf \Phi}_t{\bf F}_g\approx 0,~\forall t\neq g.
\end{equation} 
	Both design goals can be addressed by constructing ${\bf F}_g$ via the steering vector with the quantized angle-pairs covering the intended angular support (i.e., ${\left(\lambda_m^x,\lambda_{n}^y\right)\in\mathcal{A}_g}$) and excluding the other groups angular support (i.e., ${\left(\lambda_m^x,\lambda_{n}^y\right)\notin\mathcal{A}_t}$, ${\forall t\ne g}$).
The corresponding quantized angle-pairs are found as follows:
\begin{equation}\label{eq_AOD_quantized_angles}
	\left(\hat{\lambda}_{m_g}^x,\hat{\lambda}_{n_g}^y\right)
	\hspace{0.5ex}
	\Bigg|
		\hspace{0.5ex}
		\begin{array}{*{20}{l}}
			\gamma_x\in\bm{\lambda}_m^x,
			\gamma_y\in\bm{\lambda}_n^y,
			\left(\gamma_x,\gamma_y\right)\in\mathcal{A}_g,
			\\
			\left(\gamma_x,\gamma_y\right)\notin\mathcal{A}_t,\forall t\neq g,
		\end{array}
	\Bigg.
\end{equation}
where ${\boldsymbol{\lambda }}^{x}_m = \big[ {\lambda _{m}^x - \frac{1}{{{M_{x}}}},\lambda _{m}^x + \frac{1}{{{M_{x}}}}} \big]$ is the boundary of $\lambda _{m}^x$,
and 
${\boldsymbol{\lambda }}_{n}^y = \big[ {\lambda _{n}^y - \frac{1}{{{M_{y}}}},\lambda _{n}^y + \frac{1}{{{M_{y}}}}} \big]$ is the boundary of $\lambda _{n}^y$.
As shown proven in \cite[eq. (16)]{ASIL_ABHP_Access}, the approximate zero condition given in \eqref{eq_approx_zero} can be satisfied by building the analog RF beamformer via the steering vector with the corresponding quantized angle-pairs found in \eqref{eq_AOD_quantized_angles}. Thus, as the antenna array size increases, one can derive the following limit expression:
\begin{equation}\label{eq_approx_zero_proof}
	\lim\limits_{M\to \infty} 
	{\bf \Phi}_t{\bf e}\big(\hat{\lambda}_{m_g}^x,\hat{\lambda}_{n_g}^y\big)
	=
	{\bf 0},
	~
	\forall t\ne g.
\end{equation}
By combining \eqref{eq_approx_zero}, \eqref{eq_AOD_quantized_angles}, and \eqref{eq_approx_zero_proof}, it is shown that when the analog RF beamformer is designed with the corresponding quantized angle-pairs, the inter-group interference can be suppressed.
	Hence, after finding $N_{RF,g}$ quantized angle-pairs covering $\mathcal{A}_g$ and excluding $\mathcal{A}_t$ with $t\neq g$, the RF precoder for the $g^{th}$ group is constructed as follows:
\begin{equation}\label{eq_F_g}
	{\bf F}_g=\Big[
		{\bf e}\big(\hat{\lambda}_{m_1}^x,\hat{\lambda}_{n_1}^y\big),
		\cdots,
		{\bf e}\big(\hat{\lambda}_{m_{N_{RF,g}}}^x,\hat{\lambda}_{n_{N_{RF,g}}}^y\big)
	\Big].
\end{equation}
Here, $N_{RF,g}$ also represents the number of orthogonal beams generated for the $g^{th}$ group.

Finally, the analog RF beamformer ${\bf F}$ satisfying the CM constraint (i.e., $C_5$ given in \eqref{eq_obj_1}) is obtained by substituting \eqref{eq_F_g} into \eqref{eq_F}.
	Furthermore, it is important to remark that the analog RF beamformer is a unitary matrix (i.e., ${\bf F}^H {\bf F} = {\bf I}_{N_{RF}}$).

\begin{figure}[!t]
	\centering
	\subfigure[$M=16$ antennas.]
	{\includegraphics[width=\columnwidth]{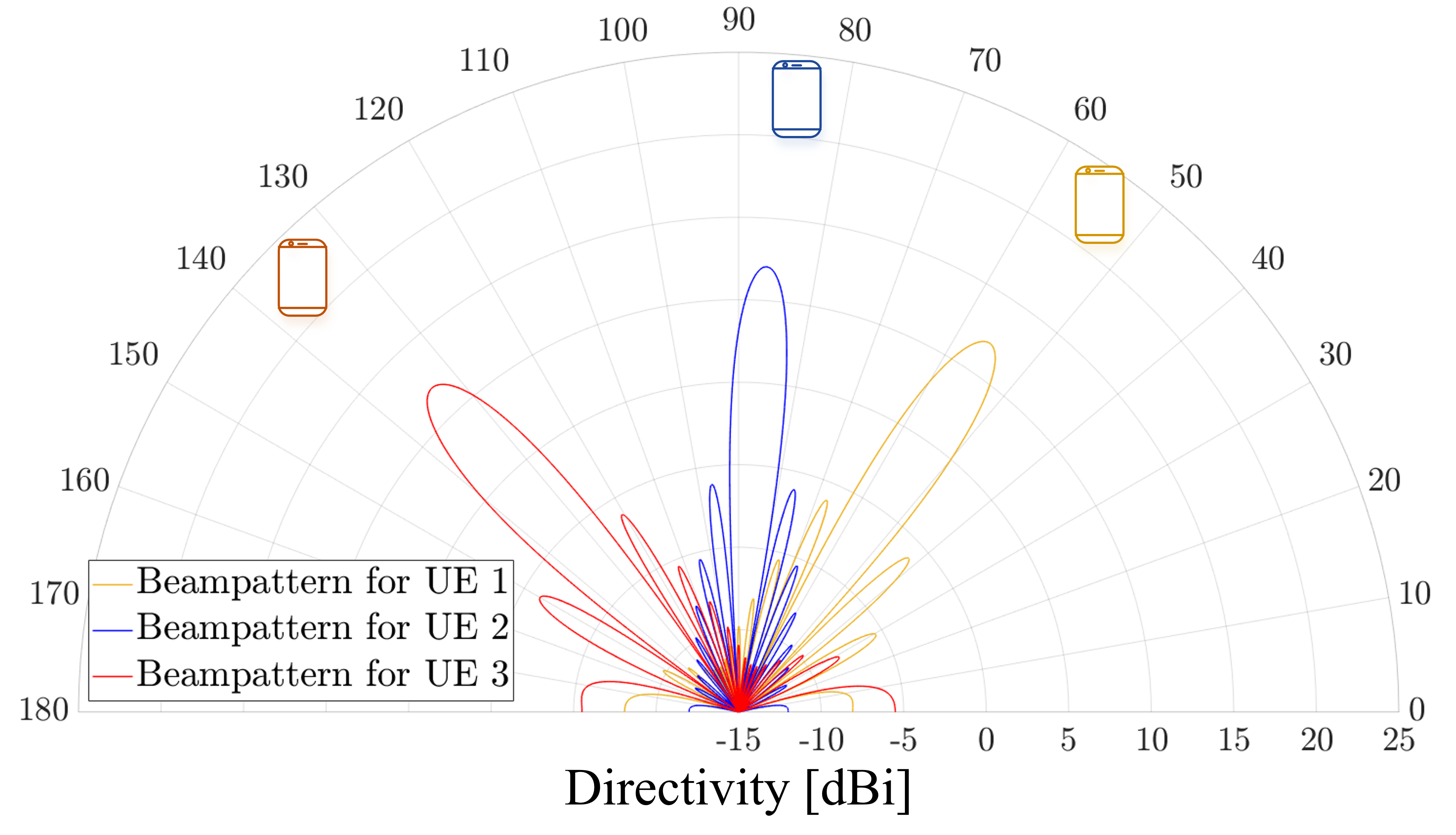}
		\label{fig_2a}}
	\subfigure[$M=64$ antennas.]
	{\includegraphics[width=\columnwidth]{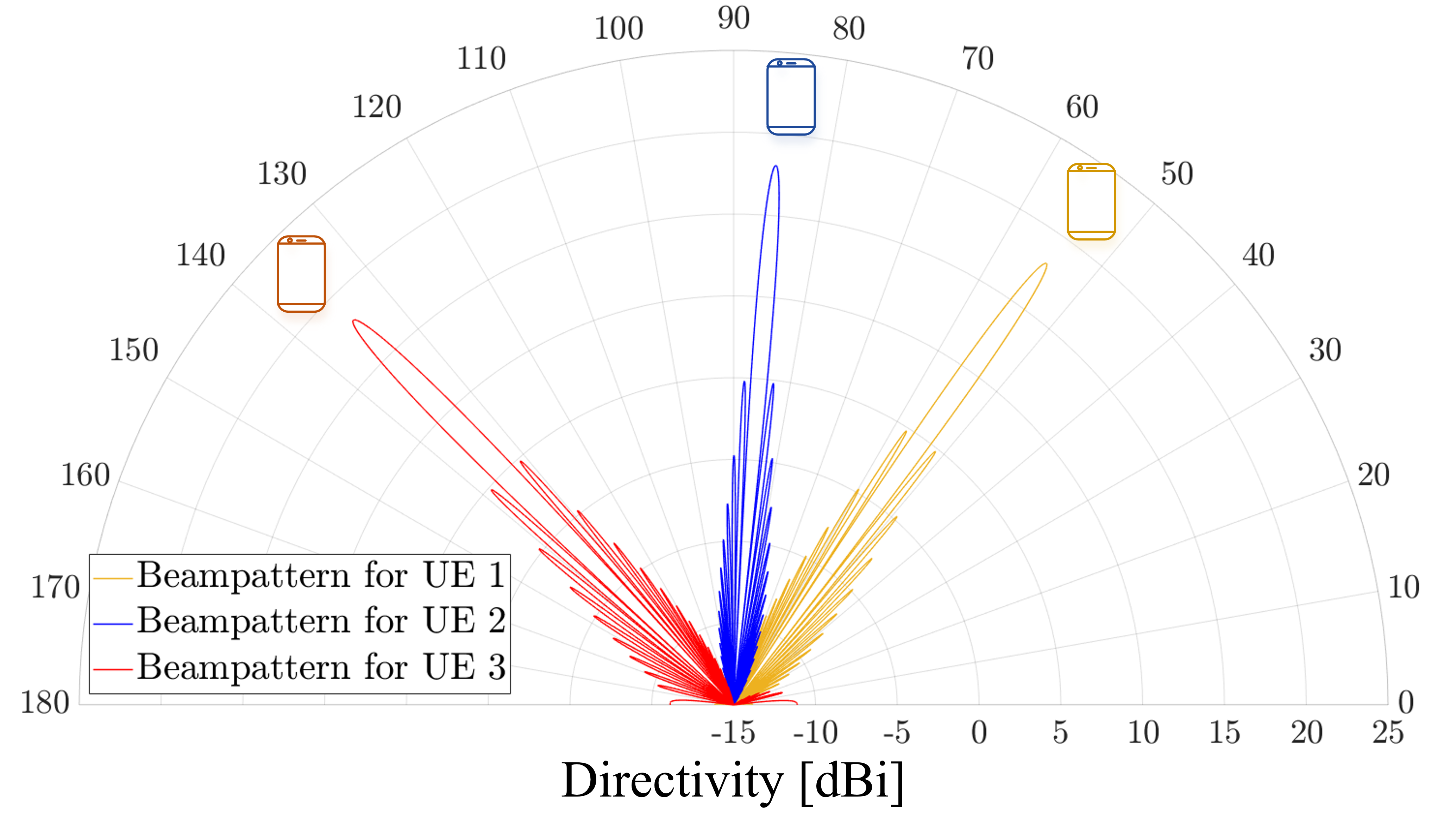}
		\label{fig_2b}}
	\caption{2D azimuth beampatterns with analog RF beamforming for each UE.
	}
	\label{fig_2}
\end{figure}

\textbf{Example 1:}
	Fig. \ref{fig_2} illustrates 2D azimuth beampatterns, where there are only $K=3$ UEs clustered in $G=3$ groups. We consider the mean AAoD of $\psi_1=55^\circ$, $\psi_2=85^\circ$, $\psi_3=135^\circ$, and the AAoD spread of $\delta^\psi_1=\delta^\psi_2=\delta^\psi_3=2^\circ$.
	For the visualization perspective, we here consider the utilization of ULA with $M=16$ and $M=64$ antennas in Fig. \ref{fig_2a} and Fig. \ref{fig_2b}, respectively.
By substituting the given angle parameters into \eqref{eq_AOD_quantized_angles}, a single beam is generated for each UE in both scenarios. In other words, by substituting \eqref{eq_steering_vec}, \eqref{eq_AOD_quantized_angles}, and \eqref{eq_F_g} into \eqref{eq_F}, the analog RF beamformer is obtained as ${\bf F}=\left[{\bf f}_1,{\bf f}_2,{\bf f}_3\right]\in\mathbb{C}^{M\times 3}$ with ${\bf f}_k={\bf e}\big(\hat{\lambda}_{m_k}^x,\hat{\lambda}_{n_k}^y\big)\in\mathbb{C}^M$ as the RF beamformer vector for the $k^{th}$ UE.
	When the number of antennas is increased from $M=16$ to $M=64$, the array directivity improves from $12$ dBi to $18$ dBi by means of higher beamforming gain. 
Although the main beam directions are orthogonal, the presence of considerable minor lobes might cause the inter-group interference. As the array size increases, we observe that the minor lobes towards undesired UE direction get smaller, which helps to mitigate the inter-group interference as indicated in \eqref{eq_approx_zero_proof}.



\subsection{Digital BB Precoder}
After deriving the analog RF beamformer ${\bf F}$, we here develop the optimal digital BB precoder ${\bf B}=\left[{\bf b}_1,\cdots,{\bf b}_K\right]$.
Hence, the $\alpha$HP objective function given in \eqref{eq_obj_1}  is reformulated accordingly:
\begin{equation}\label{eq_obj_2_BB}
	\begin{aligned}
		\max_{{\bf b}_1,\cdots,{\bf b}_K}
		&
		\sum_{k=1}^{K}
		\mathcal{U}_{\alpha}\left(R_k\right),\\
		\textrm{s.t. }
		&
		C_1:~ 
		R_k =\log_2\left( 1 + \textrm{SINR}_k \right),\\
		&
		C_2:~ 
		\textrm{SINR}_{k}
		\hspace{-0.35ex}
		=
		\frac
		{\left|{\bf h}_k^T{\bf Fb}_k\right|^2}
		{\sum_{u\ne k}^{K}\left|{\bf h}_k^T{\bf Fb}_u\right|^2 
			+
			\sigma_n^2}
		\ge
		\mu_k,\\
		&
		C_3:~ 
		\mathbb{E}\big\{\big\|{\bf s}\big\|_2^2\big\}=\sum_{k=1}^{K}{\bf b}_k^H {\bf b}_k=\sum_{k=1}^{K}p_k\le P_T,\\
		&
		C_4:~
		p_k\ge 0, \forall k,
	\end{aligned}
\end{equation}
where $\mu_1,\cdots,\mu_K$ is the set of given SINR targets maximizing the objective function \cite{yu2007transmitter}.

\textbf{Proposition 1:} The optimal digital BB precoder vector for the $k^{th}$ UE is derived as:
\begin{equation}\label{eq_BB_vector}
	{\bf{b}}_k
=
	\underbrace{\sqrt{p_k}}_{
		{\textrm{Power}}
	}
	\underbrace{
		\frac{
			\left(
				{\bf I}_{N_{RF}} + \sum_{u=1}^K \frac{\beta_u}{\sigma^2_n} {\bf A}_u
			\right)^{-1}
			{\bf F}^H {\bf h}_k^*
			}{
			\left\|
				\left(
				{\bf I}_{N_{RF}} + \sum_{u=1}^K \frac{\beta_u}{\sigma^2_n} {\bf A}_u
				\right)^{-1}
				{\bf F}^H {\bf h}_k^*
			\right\|
		}
		}_{
		\textrm{Beamforming Direction}
	}.
\end{equation}
where 
${\bf{A}}_k = {\bf F}^H {\bf h}_k^*{\bf h}_k^T{\bf F}$ and ${\sum_{k=1}^{K}p_k=\sum_{k=1}^{K}\beta_k=P_T}$. Here, $p_k$ and $\beta_k$ are parameters to be optimized, however, their closed-form solutions are not available due to the entangled optimization parameters.

\textbf{Proof:}
Please see Appendix \ref{app_NOBF_A}.

Although we obtain the optimal parameterized digital BB precoder in a closed-form, it is necessary to jointly optimize $p_k$ and $\beta_k$ for $k=1,\cdots,K$. Thus, we apply nature-inspired intelligent algorithms to optimize all $2K$ parameters.
Particularly, $N_S$ search agents are randomly placed in the $2K$ dimensional optimization space. During $Q$ iterations, the search agents communicate and compete with each other for moving towards the optimal solutions. The movement of each search agent generally depends on two major factors: (i) deterministic (i.e., the experience on the observed best solutions), (ii) stochastic (i.e., the tendency of random movement).
	
First, we define the normalized optimization parameters 
	$\hat{p}_k\in\left[0,1\right]$
	and
	$\hat{\beta}_k\in\left[0,1\right]$
with the normalization scalars of 
	$\varepsilon_1=\frac{P_T}{\sum_{k=1}^{K}\hat{p}_k}$ 
and 
	$\varepsilon_2=\frac{P_T}{\sum_{k=1}^{K}\hat{\beta}_k}$, respectively.
Afterwards, we present the $i^{th}$ search agent in the $q^{th}$ iteration as follows:
\begin{equation}\label{eq_search_agent}
	{\bf a}_{i}^{(q)}
	=
	\left[
	\hat{p}_{1,i}^{(q)},
	\cdots,
	\hat{p}_{K,i}^{(q)},
	\hat{\beta}_{1,i}^{(q)},
	\cdots,
	\hat{\beta}_{K,i}^{(q)}
	\right]^T
	\in
	\mathbb{R}^{2K},
\end{equation}
where $i=1,\cdots,N_S$ and $q=1,\cdots,Q$.
The corresponding normalization scalar is obtained as:
\begin{equation}\label{eq_normalization}
	\bm{\varepsilon}_{i,q}
	=
	\left[
	\hspace{-0.5ex}
	\begin{array}{*{20}{c}}
		{\varepsilon_{1,i}^{(q)}}\\
		{\varepsilon_{2,i}^{(q)}}
	\end{array}
	\hspace{-0.5ex}
	\right]
	=
	\left[
	\frac{P_T}{\sum_{k=1}^{K}\hat{p}_{k,i}^{(q)}},
	\frac{P_T}{\sum_{k=1}^{K}\hat{\beta}_{k,i}^{(q)}}
	\right]^T.
\end{equation}
By substituting \eqref{eq_search_agent} and \eqref{eq_normalization} into \eqref{eq_BB_vector},
the digital BB precoder vector is reformulated as a function of the search agent and normalization scalar as follows: 
\begin{equation}\label{eq_BB_vector_2}
	{\bf{b}}_k\big({\bf a},\bm{\varepsilon}\big)
	\hspace{-0.5ex}=\hspace{-0.5ex}
	\underbrace{
		\sqrt{\varepsilon_1 \hat{p}_k}
	}_{
		{\textrm{Power}}
	}
	\underbrace{
		\frac{
			\left(\hspace{-0.25ex}
			{\bf I}_{N_{RF}} \hspace{-0.75ex}+\hspace{-0.5ex} \sum_{u=1}^K\hspace{-0.5ex} \frac{\varepsilon_2\hat{\beta}_u}{\sigma^2_n} {\bf A}_u\hspace{-0.25ex}
			\right)^{\hspace{-0.5ex}-1}
			\hspace{-0.5ex}{\bf F}^H {\bf h}_k^*
		}{
			\left\|
			\left(\hspace{-0.25ex}
			{\bf I}_{N_{RF}} \hspace{-0.75ex}+\hspace{-0.5ex} \sum_{u=1}^K\hspace{-0.5ex} \frac{\varepsilon_2\hat{\beta}_u}{\sigma^2_n} {\bf A}_u\hspace{-0.25ex}
			\right)^{\hspace{-0.5ex}-1}
			\hspace{-0.5ex}{\bf F}^H {\bf h}_k^*
			\right\|
		}
	}_{
		\textrm{Beamforming Direction}
	}.
\end{equation}
By combining \eqref{eq_search_agent}, \eqref{eq_normalization}, and \eqref{eq_BB_vector_2}, the objective function defined in \eqref{eq_obj_2_BB} becomes equivalent to:
\begin{equation}\label{eq_obj_3_BB}
	\begin{aligned}
		\max_{{\bf a}}
		&
		\sum_{k=1}^{K}
		\mathcal{U}_{\alpha}\left(R_k\big({\bf a},\bm{\varepsilon}\big)\right),\\
		\textrm{s.t. }
		&
		C_1:~ 
		R_k\big({\bf a},\bm{\varepsilon}\big) =\log_2\left( 1 + \textrm{SINR}_k\big({\bf a},\bm{\varepsilon}\big) \right),\\
		&
		C_2:~ 
		\textrm{SINR}_{k}
		\hspace{-.25ex}
		\big({\bf a},\bm{\varepsilon}\big)
		\hspace{-.5ex}
		=
		\hspace{-.5ex}
		\frac
		{\left|{\bf h}_k^T{\bf Fb}_k\big({\bf a},\bm{\varepsilon}\big)\right|^2}
		{\sum_{u\ne k}^{K}\hspace{-.25ex}\left|{\bf h}_k^T{\bf Fb}_u\big({\bf a},\bm{\varepsilon}\big)\right|^2 
			\hspace{-.5ex}+\hspace{-.25ex}
			\sigma_n^2},\\
		&
		C_6:~
		{\bf a}
		=
		\left[
		\hat{p}_{1},
		\cdots,
		\hat{p}_{K,},
		\hat{\beta}_{1},
		\cdots,
		\hat{\beta}_{K}
		\right]^T
		\in\left[0,1\right], 
		\\
		&
		C_7:~
		\bm{\varepsilon}
		=
		\left[
		\frac{P_T}{\sum_{k=1}^{K}\hat{p}_k},
		\frac{P_T}{\sum_{k=1}^{K}\hat{\beta}_k}
		\right]^T,
	\end{aligned}
\end{equation}
which aims to maximize the $\alpha$-fairness objective by finding the optimal search agent (i.e., the optimal set of $p_k$ and $\beta_k$).
Here, $C_6$ and $C_7$ refer to the total transmit power constraint. 
Through iterations, we move each search agent from ${\bf a}_{i}^{(q)}$ to ${\bf a}_{i}^{(q+1)}$  with the aim of objective value maximization.
After $Q$ iterations, the best search agent is found as:
\begin{equation}\label{eq_a_best}
	{\bf a}_{\textrm{best}}
	=
	\argmax\limits_{{\bf a}_{i}^{(q)}, 
		\forall i,q
	}
	\sum_{k=1}^{K}
	\mathcal{U}_{\alpha}\left(R_k\big(
	{\bf a}_{i}^{(q)},
	\bm{\varepsilon}_{i}^{(q)}\big)
	\right).
\end{equation}
Now, we propose to apply five nature-inspired intelligent algorithms to describe how search agents move over iterations. 

\subsubsection{PSO-$\alpha$HP}
Particle swarm optimization (PSO) is classified as a swarm intelligence algorithm, which is inspired by the swarm behavior of animals \cite{PSO_AI}.
	In the proposed PSO-$\alpha$HP technique, the position of each particle (i.e., search agent) is updated as:
\begin{equation}\label{eq_PSO_clipping}
	{\bf a}_{i}^{(q+1)} =\textrm{clip}
	\big(
	{\bf a}_{i}^{(q)} + \bm{\Delta}_{i}^{(q)},
	\left[0,1\right]
	\big),
\end{equation}
where $\bm{\Delta}_{i}^{(q)}\in\mathbb{R}^{2K}$ denotes the velocity of the $i^{th}$ particle in the $q^{th}$ iteration. 
Also, we define a clipping function as $\textrm{clip}\left(x,\left[a,b\right]\right)=\max\left(a,\min\left(x,b\right)\right)$
to keep the normalized optimization parameters within their predefined ranges.

Throughout iterations, when a particle finds a better position achieving higher objective value than any of its previous positions, it is recorded as its current best position.
In the $q^{th}$ iteration, the velocity of each particle is calculated based on three factors: (i) its current personal best position 
	${\bf a}_{\textrm{best},i}^{(q)}$, 
(ii) the current global best position among all particles 
	${\bf a}_ {\textrm{gbest}}^{(q)}$, 
(iii) randomness.
	Thus, the velocity in the $(q+1)^{th}$ iteration is updated as follows:
\begin{equation}\label{eq_PSO_velocity}
	\begin{aligned}
		\bm{\Delta}_{i}^{(q+1)}
		&=
		\textrm{clip}\Big(\Big.
		\kappa_{q}^{\textrm{PSO}}
		\bm{\Delta}_{i}^{(q)}
			+
		{\bf w}_1^{\textrm{PSO}}
			\hspace{-0.25ex}\odot\hspace{-0.25ex}
		\big(
		{\bf a}_{\textrm{gbest}}^{(q)}
			-
		{\bf a}_{i}^{(q)}
		\big)\\
		&
			+
		{\bf w}_2^{\textrm{PSO}}
			\hspace{-0.25ex}\odot\hspace{-0.25ex}
		\big(
		{\bf a}_{\textrm{best},i}^{(q)}
			-
		{\bf a}_{i}^{(q)}
		\big),
		\Big.
			\left[
				\kappa_{\Delta,\min}^{\textrm{PSO}},
				\kappa_{\Delta,\max}^{\textrm{PSO}}
			\right]
		\Big),
	\end{aligned}
\end{equation}
where
	$\kappa_{q}^{\textrm{PSO}}
	=
	1-\frac{q-1}{Q}$ is a decreasing inertia parameter for enhancing the exploitation towards the latest iterations,
	$\kappa_{\Delta,\min}^{\textrm{PSO}}$
		and
	$\kappa_{\Delta,\max}^{\textrm{PSO}}$
		are respectively the minimum and maximum velocity parameters
	 \cite{PSO_book}.
Also,
	${\bf w}_1^{\textrm{PSO}}\in\mathbb{R}^{2K}$ 
		and
	${\bf w}_2^{\textrm{PSO}}\in\mathbb{R}^{2K}$ 
		are random vectors with uniformly distributed entries over  $\left[0,2\right]$, 
	which determine the tendency of moving towards the current global and personal best positions to explore the $2K$ dimensional optimization space.
{
	Particularly, the current global and personal best positions in the ${(q+1)}^{th}$ iteration are  respectively found as:}
\begin{equation}\label{eq_PSO_gbest}
	\begin{aligned}
		{\bf a}_{\textrm{gbest}}^{(q+1)}
			=
		\argmax\limits_{
			{{\bf a}_{i}^{(q')},}
			{\forall i,\forall q'\le q}
		} 
		\hspace{1ex}
		\sum_{k=1}^{K}
		\mathcal{U}_{\alpha}\left(R_k\big(
		{\bf a}_{i}^{(q')},
		\bm{\varepsilon}_{i}^{(q')}\big)
		\right),
	\end{aligned}
\end{equation}
\begin{equation}\label{eq_PSO_pbest}
	\begin{aligned}
	\hspace{-2.2ex}{\bf a}_{\textrm{best},i}^{(q+1)}
		=
	\argmax\limits_{
		{{\bf a}_{i}^{(q')},}
		{\forall q'\le q}
	} 
	\hspace{1ex}
	\sum_{k=1}^{K}
	\mathcal{U}_{\alpha}\left(R_k\big(
	{\bf a}_{i}^{(q')},
	\bm{\varepsilon}_{i}^{(q')}\big)
	\right).
	\end{aligned}
\end{equation}
After $Q$ iterations, PSO-$\alpha$HP finally finds the best position as ${\bf a}_{\textrm{best}} = {\bf a}_{\textrm{gbest}}^{(Q)}$, which is substituted into \eqref{eq_normalization} and \eqref{eq_BB_vector_2} to construct the digital BB precoder. 
{Algorithm \ref{alg_PSO} summarizes the proposed PSO-$\alpha$HP technique for the optimal digital BB precoder design.}
	
\begin{algorithm}[!t]
	\caption{PSO-$\alpha$HP for optimal digital BB precoder}
	\label{alg_PSO}
	\begin{algorithmic}[1]
		\renewcommand{\algorithmicrequire}{\bf{Input:}}
		\renewcommand{\algorithmicensure}{\bf{Output:}}
		\Require 
		$\bm{\mathcal{H}}$, $K$, $P_T$, $Q$, $N_S$, $\kappa_{\Delta,\max}^{\textrm{PSO}}$, $\kappa_{\Delta,\min}^{\textrm{PSO}}$.
		\State Randomly initialize $N_S$ particles ${\bf a}_{i}^{(0)}$, $\forall i=1:N_S$.
		\State Set the initial personal best solution as ${\bf a}_{\textrm{best},i}^{(0)}={\bf a}_{i}^{(0)}$.
		\State Find the initial global best position ${\bf a}_{\textrm{gbest}}^{(0)}$ via \eqref{eq_PSO_gbest}.
		\State {\textbf{for}} $q=1:Q$, {\textbf{do}}
		\State ~~ {\textbf{for}} $i=1:N_S$, {\textbf{do}}
		\State ~~~ Calculate velocity $\bm{\Delta}_{i}^{(q)}$ via \eqref{eq_PSO_velocity}.
		\State ~~~ Update the particle position ${\bf a}_{i}^{(q)}$ via \eqref{eq_PSO_clipping}.
		\State ~~~ Find the current personal best position ${\bf a}_{\textrm{best},i}^{(q)}$ via \eqref{eq_PSO_pbest}.
		\State ~~ {\textbf{end for}}
		\State ~~ Find the current global best position ${\bf a}_{\textrm{gbest}}^{(q)}$ via \eqref{eq_PSO_gbest}.
		\State {\textbf{end for}}
		\Ensure ${\bf a}_{\textrm{best}} = {\bf a}_{\textrm{gbest}}^{(Q)}$.
	\end{algorithmic}
\end{algorithm}
	
\subsubsection{GWO-$\alpha$HP}
Grey wolf optimization (GWO) mathematically implements the leadership hierarchy and hunting mechanisms of the grey wolves, which are apex predators at the top of food chain \cite{GWO_first_paper}.
Unlike PSO, instead of moving towards its personal best, each wolf (e.g., search agent) follows three best wolves classified as alpha, beta, and delta. Specifically, the alpha wolf is considered as the group leader, who is located at the global best position with the highest objective value. Similarly, the beta wolf and delta wolf represent the second and third global best positions, respectively. Based on the variation in the objective values, the alpha, beta, and delta wolves might be updated in every iteration. Hence, there is a stringent competition among the three best wolves, which eventually benefits the increase of objective value. In the proposed GWO-$\alpha$HP technique, 
the position of each wolf in the $(q+1)^{th}$ iteration is found as\footnote{Similar to \eqref{eq_PSO_clipping}, we always apply the clipping function in order to keep the search agents in $\left[0,1\right]$. However, it is omitted for the brevity, when the search agent moves are expressed in \eqref{eq_GWO_move}, \eqref{eq_ACO_move}, \eqref{eq_CS_move_1}, \eqref{eq_CS_move_2} and \eqref{eq_FA_move}.}:
\begin{equation}\label{eq_GWO_move}
	\begin{aligned}
		{\bf a}_{i}^{(q+1)}
		&
			\hspace{-0.5ex}=\hspace{-0.5ex}
		\sum_{j=1}^{3}
		\frac{
		{\bf a}_{\textrm{wolf},j}^{(q)}
			\hspace{-.25ex}-\hspace{-0.25ex}
		{\bf w}_{j,1,q}^{\textrm{GWO}}
			\hspace{-0.5ex}\odot\hspace{-0.5ex}
		\left|
			{\bf w}_{j,2}^{\textrm{GWO}}
				\hspace{-0.25ex}\odot\hspace{-0.25ex}
			{\bf a}_{\textrm{wolf},j}^{(q)} 
				\hspace{-0.25ex}-\hspace{-0.25ex}
			{\bf a}_{i}^{(q)}\right|
		}{
			3
		},\hspace{-2ex}
	\end{aligned}
\end{equation}
where ${\bf a}_{\textrm{wolf},j}^{(q)}$ denotes the $j^{th}$ best wolf in the $q^{th}$ iteration (i.e., $j=1$ for alpha, $j=2$ for beta, $j=3$ for delta),
	${\bf w}_{j,1,q}^{\textrm{GWO}}$ 
		and 
	${\bf w}_{j,2}^{\textrm{GWO}}$
		are random vectors with the uniformly distributed entries over
	$\left[-\kappa_{q}^{\textrm{GWO}},+\kappa_{q}^{\textrm{GWO}}\right]$
		 and
 	$\left[0,2\right]$, respectively.
Here, $\kappa_{q}^{\textrm{GWO}}=2-\frac{2q-2}{Q}$ reduces from $2$ to $0$ for switching from exploration to exploitation, respectively \cite{GWO_first_paper}.
	By using \eqref{eq_a_best}, the alpha wolf in the $Q^{th}$ iteration is selected as the best search agent as ${\bf a}_{\textrm{best}} = {\bf a}_{\textrm{wolf},1}^{(Q)}$.
{The proposed GWO-$\alpha$HP technique is presented in Algorithm \ref{alg_GWO}.}
\begin{algorithm}[!t]
	\caption{GWO-$\alpha$HP for optimal digital BB precoder}
	\label{alg_GWO}
	\begin{algorithmic}[1]
		\renewcommand{\algorithmicrequire}{\bf{Input:}}
		\renewcommand{\algorithmicensure}{\bf{Output:}}
		\Require 
		$\bm{\mathcal{H}}$, $K$, $P_T$, $Q$, $N_S$.
		\State Randomly initialize $N_S$ wolves ${\bf a}_{i}^{(0)}$, $\forall i=1:N_S$.
		\State Find the initial three best position ${\bf a}_{\textrm{wolf},j}^{(0)}$, $\forall j=1:3$.
		\State {\textbf{for}} $q=1:Q$, {\textbf{do}}
		\State ~~ {\textbf{for}} $i=1:N_S$, {\textbf{do}}
		\State ~~~~ Update the wolf position ${\bf a}_{i}^{(q)}$ via \eqref{eq_GWO_move}.
		\State ~~ {\textbf{end for}}
		\State ~~ Find the current three best position ${\bf a}_{\textrm{wolf},j}^{(q)}$, $\forall j=1:3$.
		\State {\textbf{end for}}
		\Ensure ${\bf a}_{\textrm{best}} = {\bf a}_{\textrm{wolf},1}^{(Q)}$.
	\end{algorithmic}
\end{algorithm}
 	
\subsubsection{ACO-$\alpha$HP}

Ant colony optimization (ACO) mimics the communication mechanism among the ants while searching the food sources around their nest \cite{ACO_cont}. 
	In an ant colony, each ant (e.g., search agent) regularly seeks to find the food sources in various directions. 
After discovering a food source, the ant marks the corresponding path with an evaporating chemical called \textit{pheromone} while returning to the nest with food.
	The amount of pheromone deposited by each ant varies based on the food quantity/quality, which is considered as the objective value in the perspective of optimization.
By the time, the ants prefer moving in the direction of routes with dense pheromone levels instead of entirely random search.
	This indirect communication mechanism motivates the development of ACO algorithm. Although ACO is originally proposed for the combinatorial optimization problems, it is also successfully adopted in the continuous optimization problems \cite{ACO_cont}.
	
For the mathematical foundation of the proposed ACO-$\alpha$HP technique, $N_S$ ants discover a food source position ${\bf a}_{i}^{(q)}$ in every iteration. Afterwards, only $\kappa_1^{\textrm{ACO}}$ of them with the highest objective values are archived. 
	Specifically, 
	${\bf a}_{\textrm{ant},j}^{(q)}$ represents the
	$j^{th}$ best food source position with $j=1,\cdots,\kappa_1^{\textrm{ACO}}$ discovered by the ant colony after the $q^{th}$ iteration.
By applying the Gaussian kernel, the pheromone level for the $j^{th}$ food source
is defined by \cite{ACO_cont}:
\begin{equation}\label{eq_ACO_kernel}
	\Upsilon_{j}^{\textrm{ACO}}
	=
	\frac{
		1}{
		\kappa_1^{\textrm{ACO}} 
		\kappa_2^{\textrm{ACO}}
		\sqrt{2\pi}
	}
	e^{
		-\frac{1}{2}
		\frac{
			\left(j-1\right)^2
		}{
			\left(
			\kappa_1^{\textrm{ACO}} 
			\kappa_2^{\textrm{ACO}}
			\right)^2
		}
	},
\end{equation}
where $\kappa_2^{\textrm{ACO}}$ is a parameter specifying the pheromone level difference among the food sources.
	Based on the pheromone levels of a set of $\kappa_1^{\textrm{ACO}}$  food sources, an ant prefers moving towards the $j^{th}$ food source with the probability of:
\begin{equation}\label{eq_ACO_prob}
	\mathcal{P}_j^{\textrm{ACO}}
	=
	\frac{\Upsilon_{j}^{\textrm{ACO}}}
	{\sum_{c=1}^{\kappa_1^{\textrm{ACO}}}~\Upsilon_{c}^{\textrm{ACO}}}\in\left[0,1\right].
\end{equation}
By using \eqref{eq_ACO_kernel} and \eqref{eq_ACO_prob},
	one can show that when $\kappa_2^{\textrm{ACO}}$ is small, the best food source position is highly preferred.
Otherwise, for the larger values of $\kappa_2^{\textrm{ACO}}$, the probability of choosing each food source position becomes almost equivalent.
	Afterwards, the $v^{th}$ dimension of the $i^{th}$ ant position is calculated in the ${(q+1)^{th}}$ iteration as follows:
\begin{equation}\label{eq_ACO_move}
	\begin{aligned}
		{\bf a}_{i}^{(q+1)}
		\hspace{-0.5ex}
		\left[v\right]
		\hspace{-0.25ex}
		=
		\hspace{-0.25ex}
		{\bf a}_{\textrm{ant},j}^{(q)}
		\left[v\right]
		+
		w^{\textrm{ACO}}
		\kappa_{j,q,v}^{\textrm{ACO}},
	\end{aligned}
\end{equation}
where 
$
	v =1,\cdots,2K
$
	represents the dimension index of an ant position,
${\bf a}_{\textrm{ant},j}^{(q)}\left[v\right]$ denotes the randomly chosen $j^{th}$ food source position for the corresponding dimension via \eqref{eq_ACO_prob},
	${w}^{\textrm{ACO}}$ 
	is a random scalar following the distribution of $\mathcal{N}\left(0,1\right)$,
	$\kappa_{j,q,v}^{\textrm{ACO}}
		=
		\sum_{c=1}^{\kappa_1^{\textrm{ACO}}}
		{\big|
			{\bf a}_{\textrm{ant},j}^{(q)}\left[v\right]
			-
			{\bf a}_{\textrm{ant},c}^{(q)}\left[v\right]
		\big|}
				/
		\big({\kappa_1^{\textrm{ACO}}-1}\big)
	$ expresses the average distance from the chosen food source position to the others.
According to the new objective values, $\kappa_1^{\textrm{ACO}}$ best food source positions are updated for the next iteration. 
	After applying this procedure for $Q$ iterations, we obtain the best position as ${\bf a}_{\textrm{best}} = {\bf a}_{\textrm{ant},1}^{(Q)}$. 
{Finally, Algorithm \ref{alg_ACO} expresses the proposed ACO-$\alpha$HP technique.}

\begin{algorithm}[!t]
	\caption{ACO-$\alpha$HP for optimal digital BB precoder}
	\label{alg_ACO}
	\begin{algorithmic}[1]
		\renewcommand{\algorithmicrequire}{\bf{Input:}}
		\renewcommand{\algorithmicensure}{\bf{Output:}}
		\Require 
		$\bm{\mathcal{H}}$, $K$, $P_T$, $Q$, $N_S$, $\kappa_1^{\textrm{ACO}}$,
		$\kappa_2^{\textrm{ACO}}$.
		\State Randomly initialize $N_S$ ants ${\bf a}_{i}^{(0)}$, $\forall i=1:N_S$.
		\State Find the initial $\kappa_1^{\textrm{ACO}}$ best position ${\bf a}_{\textrm{ant},j}^{(0)}$, $\forall j=1:\kappa_1^{\textrm{ACO}}$.
		\State Calculate $\mathcal{P}_j^{\textrm{ACO}}$ via \eqref{eq_ACO_kernel} and \eqref{eq_ACO_prob}, $\forall j=1:\kappa_1^{\textrm{ACO}}$, as the probability of choosing the $j^{th}$ best food source.
		\State {\textbf{for}} $q=1:Q$, {\textbf{do}}
		\State ~~ {\textbf{for}} $i=1:N_S$, {\textbf{do}}
		\State ~~~~ Randomly choose the $j^{th}$ best food source ${\bf a}_{\textrm{ant},j}^{(q-1)}$.
		\State ~~~~ Update the ant position ${\bf a}_{i}^{(q)}$ via \eqref{eq_ACO_move}.
		\State ~~ {\textbf{end for}}
		\State ~~ Find the current $\kappa_1^{\textrm{ACO}}$ best position ${\bf a}_{\textrm{ant},j}^{(q)}$, $\forall j=1:\kappa_1^{\textrm{ACO}}$.
		\State {\textbf{end for}}
		\Ensure ${\bf a}_{\textrm{best}} = {\bf a}_{\textrm{ant},1}^{(Q)}$.
	\end{algorithmic}
\end{algorithm}

\subsubsection{CS-$\alpha$HP}
Cuckoo search (CS) algorithm is inspired by the aggressive reproduction strategy of the cuckoo birds \cite{PSO_book}. Particularly, each cuckoo (e.g., search agent) lays its own eggs to the nests of other cuckoos. Furthermore, it might remove the eggs belonging to other cuckoos for increasing the survival probability of its own eggs towards the next generations (i.e., iterations).
		Here, the overall egg quality in a nest represents the objective value. 
	Thus, the CS algorithm includes three main steps in each iteration:
		(i)   each cuckoo randomly flies to a new nest following the Lévy flights and lays its eggs,
		(ii)  select either new or previous nest by comparing their egg quality,
		(iii) the host cuckoo can replace the suspected eggs with a probability of $\kappa_1^{\textrm{CS}}\in\left[0,1\right]$ as long as it improves the overall egg quality.

In the proposed CS-$\alpha$HP technique, we first temporarily change the $v^{th}$ dimension of the $i^{th}$ cuckoo position according to the Lévy flight as follows \cite{PSO_book}:
\begin{equation}\label{eq_CS_move_1}
	{\bf a}_{\textrm{CS}_1,i}^{(q+1)}
	\hspace{-0.5ex}\left[v\right]
	=
	{\bf a}_{i}^{(q)}
	\hspace{-0.5ex}\left[v\right]
	+
	\frac{\kappa_2^{\textrm{CS}}}{\pi}
	\frac{
		\Gamma\left(\kappa_2^{\textrm{CS}}\right)		
		\sin\left(\kappa_2^{\textrm{CS}}\frac{\pi}{2}\right)
	}{
		\left| w_1^{\textrm{CS}}\right|^{1+\kappa_2^{\textrm{CS}}}
	}
\end{equation}
where $\Gamma\left(\cdot\right)$ is the standard gamma function, 
	$\kappa_2^{\textrm{CS}}$ is a parameter for the Lévy exponent, 
	$w_1^{\textrm{CS}}
		=
	\frac{w_2^{\textrm{CS}}}{|w_3^{\textrm{CS}}|^{1/\kappa_2^{\textrm{CS}}}}$ is a random step size based on two Gaussian random variables
	$w_2^{\textrm{CS}}\sim\mathcal{N}
	\bigg(0,
			\Big[\frac{
				\Gamma\left(1+\kappa_2^{\textrm{CS}}\right)
			}{
				\kappa_2^{\textrm{CS}}
				\Gamma\left( (\kappa_2^{\textrm{CS}} + 1)/2\right)
			}
			\frac{
					\sin\left(\kappa_2^{\textrm{CS}}\pi/2\right)
				}{
					2^{(\kappa_2^{\textrm{CS}}-1)/2}
				}
		\Big]^{1/\kappa_2^{\textrm{CS}}}
	\bigg)$
	and 
	$w_3^{\textrm{CS}}\sim\mathcal{N}\left(0,1\right)$.
Afterwards,
	we either select ${\bf a}_{\textrm{CS}_1,i}^{(q+1)}$ or ${\bf a}_{i}^{(q)}$ by comparing their objective values, then it is assigned to another temporary position as 
	${\bf a}_{\textrm{CS}_2,i}^{(q+1)}$.
In the third and last step, we apply a random replacement of the suspected eggs with the probability of $\kappa_1^{\textrm{CS}}\in\left[0,1\right]$.
	Thus, the
$i^{th}$ cuckoo temporary position can be replaced with any randomly chosen $j^{th}$ cuckoo temporary position as follows \cite{PSO_book}:
\begin{equation}\label{eq_CS_move_2}
	{\bf a}_{\textrm{CS}_3,i}^{(q+1)}
	\hspace{-0.5ex}\left[v\right]
	=
	\left(1-w_4^{\textrm{CS}}\right)
	{\bf a}_{\textrm{CS}_2,i}^{(q+1)}
	\hspace{-0.5ex}\left[v\right]
	+
	w_4^{\textrm{CS}}
	{\bf a}_{\textrm{CS}_2,j}^{(q+1)}
	\hspace{-0.5ex}\left[v\right],
\end{equation}
where 
	$w_4^{\textrm{CS}}\sim\textrm{Ber}\left(\kappa_1^{CS}\right)$ is a Bernoulli random variable (i.e., $w_4^{\textrm{CS}}=1$ with the probability of $\kappa_1^{CS}$, otherwise, $w_4^{\textrm{CS}}=0$). Then, the best temporary position among
	${\bf a}_{\textrm{CS}_1,i}^{(q+1)}$,
	${\bf a}_{\textrm{CS}_2,i}^{(q+1)}$,
	 and
	${\bf a}_{\textrm{CS}_3,i}^{(q+1)}$ is transferred to the next iteration as ${\bf a}_{i}^{(q+1)}$.
Finally, when we complete $Q$ iterations, the proposed CS-$\alpha$HP finds the best position ${\bf a}_{\textrm{best}}$ in the last iteration via \eqref{eq_obj_3_BB}.
	{The proposed CS-$\alpha$HP technique is summarized in Algorithm \ref{alg_CS}.}
\begin{algorithm}[!t]
	\caption{CS-$\alpha$HP for optimal digital BB precoder}
	\label{alg_CS}
	\begin{algorithmic}[1]
		\renewcommand{\algorithmicrequire}{\bf{Input:}}
		\renewcommand{\algorithmicensure}{\bf{Output:}}
		\Require 
		$\bm{\mathcal{H}}$, $K$, $P_T$, $Q$, $N_S$, $\kappa_1^{\textrm{CS}}$,
		$\kappa_2^{\textrm{CS}}$.
		\State Randomly initialize $N_S$ cuckoos ${\bf a}_{i}^{(0)}$, $\forall i=1:N_S$.
		\State {\textbf{for}} $q=1:Q$, {\textbf{do}}
		\State ~~ {\textbf{for}} $i=1:N_S$, {\textbf{do}}
		\State ~~~~ Calculate the first temporary position $	{\bf a}_{\textrm{CS}_1,i}^{(q)}$ via \eqref{eq_CS_move_1}.
		\State ~~~~ Select ${\bf a}_{\textrm{CS}_1,i}^{(q)}$ or ${\bf a}_{i}^{(q-1)}$ based on objective value, then  \Statex ~~~~ assign it to the second temporary position ${\bf a}_{\textrm{CS}_2,i}^{(q)}$.
		\State ~~ {\textbf{end for}}
		\State ~~ {\textbf{for}} $i=1:N_S$, {\textbf{do}}
		\State ~~~~ Randomly choose a cuckoo's second temporary
		\Statex ~~~~ position as $	{\bf a}_{\textrm{CS}_2,j}^{(q)}$ with $j\in\left\{1,\cdots,N_S\right\}$.
		\State ~~~~ Calculate the third temporary position $	{\bf a}_{\textrm{CS}_3,i}^{(q)}$ via \eqref{eq_CS_move_2}.
		\State ~~~~ Update the cuckoo position ${\bf a}_{i}^{(q)}$ by selecting the best
		\Statex ~~~~ of 	
		${\bf a}_{\textrm{CS}_1,i}^{(q)}$,
		${\bf a}_{\textrm{CS}_2,i}^{(q)}$,
		and
		${\bf a}_{\textrm{CS}_3,i}^{(q)}$.
		\State ~~ {\textbf{end for}}
		\State {\textbf{end for}}
		\State Find the best cuckoo ${\bf a}_{\textrm{best}}$ via \eqref{eq_obj_3_BB}.
		\Ensure ${\bf a}_{\textrm{best}}$.
	\end{algorithmic}
\end{algorithm}

\subsubsection{FA-$\alpha$HP}
Firefly algorithm (FA) is motivated by the flashing light communication strategy, where
		each firefly (i.e., search agent) aims to attract the mating partners according to the brightness of its flashing light \cite{PSO_book}. Here, the objective value is described by the brightness.
On the other hand, 	the attractiveness of each firefly pair decays exponentially on the distance between them.
	In every iteration, each firefly finds the best mating partner among all combinations.

Unlike the previous four NI-$\alpha$HP techniques, the proposed FA-$\alpha$HP only employs  $\sqrt{N_S}$ search agents in order to have the same order of computational complexity \cite{PSO_book}. In other words,
	considering $\sqrt{N_S}$ fireflies and $\sqrt{N_S}$ mating opportunity for each firefly, we compute
$
N_S
	=
\sqrt{N_S}\times\sqrt{N_S}
$
combinations
	for 
the possible firefly positions per iteration.
Therefore, the proposed FA-$\alpha$HP first computes all possible positions for the $i^{th}$ firefly with respect to any mating partner as follows:
\begin{equation}\label{eq_FA_move}
	{\bf a}_{\textrm{FA},i\to j}^{(q+1)}
		\hspace{-0.25ex}=\hspace{-0.25ex}
	{\bf a}_{i}^{(q)}
		\hspace{-0.25ex}+\hspace{-0.15ex}
	e^{-\kappa_1^{\textrm{FA}}\Upsilon_{i,j,q}^{\textrm{FA}}}
	\hspace{-0.5ex}
	\left[
		\hspace{-0.15ex}
		{\bf a}_{j}^{(q)}
			\hspace{-0.5ex}-\hspace{-0.5ex}
		{\bf a}_{i}^{(q)}
		\hspace{-0.25ex}
	\right]
		\hspace{-0.25ex}+\hspace{-0.25ex}
	\left[\kappa_2^{\textrm{FA}}\right]^{q}
	{\bf w}^{\textrm{FA}},
\end{equation}
where
	$\forall i,j=1,\cdots,\sqrt{N_S}$ denotes the firefly index,
	$
	\Upsilon_{i,j,q}^{\textrm{FA}}
		=
	\big\|
	{\bf a}_{i}^{(q)}
	-
	{\bf a}_{j}^{(q)}
	\big\|^2
	$
indicates the distance between the corresponding firefly pairs,
	${\bf w}^{\textrm{FA}}$ is the random vector with the uniformly distributed entries over $\left[-\frac{1}{2},+\frac{1}{2}\right]$,
	$\kappa_1^{\textrm{FA}}$
		and
	$\kappa_2^{\textrm{FA}}$
		are the parameters for the light absorption and randomness, respectively \cite{PSO_book}.
For the $i^{th}$ firefly, the best position with the highest objective value among all possible ${\bf a}_{\textrm{FA},i\to j}^{(q)}$ is assigned to
	${\bf a}_{i}^{(q+1)}$.
	This procedure is applied for each firefly through $Q$ iterations, then, FA-$\alpha$HP finds the best position ${\bf a}_{\textrm{best}}$ via \eqref{eq_obj_3_BB}.
{Algorithm \ref{alg_FA} explains the proposed FA-$\alpha$HP technique for the optimal digital BB precoder design.}

\begin{algorithm}[!t]
	\caption{FA-$\alpha$HP for optimal digital BB precoder}
	\label{alg_FA}
	\begin{algorithmic}[1]
		\renewcommand{\algorithmicrequire}{\bf{Input:}}
		\renewcommand{\algorithmicensure}{\bf{Output:}}
		\Require 
		$\bm{\mathcal{H}}$, $K$, $P_T$, $Q$, $N_S$, $\kappa_1^{\textrm{FA}}$,
		$\kappa_2^{\textrm{FA}}$.
		\State Randomly initialize $\sqrt{N_S}$ fireflies ${\bf a}_{i}^{(0)}$, $\forall i=1:\sqrt{N_S}$.
		\State {\textbf{for}} $q=1:Q$, {\textbf{do}}
		\State ~~ {\textbf{for}} $i=1:\sqrt{N_S}$, {\textbf{do}}
		\State ~~~~ {\textbf{for}} $j=1:\sqrt{N_S}$, {\textbf{do}}
		\State ~~~~~~ When the $j^{th}$ firefly is the mating partner of
		\Statex ~~~~~~ the $i^{th}$ firefly, calculate ${\bf a}_{\textrm{FA},i\to j}^{(q)}$  via \eqref{eq_FA_move}.
		\State ~~~~ {\textbf{end for}}
		\State ~~ Update the firefly position ${\bf a}_{i}^{(q)}$ with the best of 
		${\bf a}_{\textrm{FA},i\to j}^{(q)}$.
		\State ~~ {\textbf{end for}}
		\State {\textbf{end for}}
		\State Find the best firefly ${\bf a}_{\textrm{best}}$ via \eqref{eq_obj_3_BB}.
		\Ensure ${\bf a}_{\textrm{best}}$.
	\end{algorithmic}
\end{algorithm}

Algorithm \ref{alg_NIaHP} summarizes the proposed NI-$\alpha$HP technique, which consecutively develops the analog RF beamformer ${\bf F}$ and digital BB precoder ${\bf B}$.
	In addition to the $\alpha$-fairness,
	NI-$\alpha$HP considerably reduces both the number of RF chains
	(i.e., from $M$ to $N_{RF}$)\footnote{By applying the transfer block design in \cite{ASIL_ABHP_Access}, NI-$\alpha$HP can be also implemented via only $K$ RF chains by keeping the exactly same performance.}.
and
the channel estimation overhead size
	(i.e., from $M\times K$ to $N_{RF}\times K$).

\begin{algorithm}[!t]
	\caption{Nature-inspired $\alpha$-fair hybrid precoding}
	\label{alg_NIaHP}
	\centering --- Analog RF Beamformer ---
	\begin{algorithmic}[1]
		\renewcommand{\algorithmicrequire}{\bf{Input:}}
		\renewcommand{\algorithmicensure}{\bf{Output:}}
		\Require 
		$M$, $G$, $\theta_g$, $\delta_g^\theta$, $\psi_g$, $\delta_g^\psi$, $\forall g=1,\cdots,G$.
		\State {\textbf{for}} $g=1:G$, {\textbf{do}}
		\State ~~ Build angular support $\mathcal{A}_g$ via \eqref{eq_AoD_Supp}.
		\State ~~ Find angle-pairs $\big(\hat{\lambda}_{m_g}^x,\hat{\lambda}_{n_g}^y\big)$ covering $\mathcal{A}_g$ via \eqref{eq_AOD_quantized_angles}.
		\State ~~ Develop RF beamformer ${\bf F}_g$ via \eqref{eq_F_g}.
		\State {\textbf{end for}}
		\Ensure ${\bf F}= \left[{\bf F}_1,\cdots,{\bf F}_G\right]$.
	\end{algorithmic}
	\centering --- Digital BB Precoder ---
	\begin{algorithmic}[1]
		\renewcommand{\algorithmicrequire}{\bf{Input:}}
		\renewcommand{\algorithmicensure}{\bf{Output:}}
		\Require 
		$\bm{\mathcal{H}}$, $K$, $P_T$.
		\State { Optimize ${\bf a}
		=
		\big[
		\hat{p}_{1},
		\cdots,
		\hat{p}_{K},
		\hat{\beta}_{1},
		\cdots,
		\hat{\beta}_{K}
		\big]^T$ 
			by applying
		PSO-$\alpha$HP, 
		GWO-$\alpha$HP, 
		ACO-$\alpha$HP, 
		CS-$\alpha$HP, or
		FA-$\alpha$HP.
			via
		Algorithm 
		\ref{alg_PSO}, 
		\ref{alg_GWO}, 
		\ref{alg_ACO}, 
		\ref{alg_CS}, 
		or
		\ref{alg_FA}, respectively.}
		\State  Calculate scalars
		$\varepsilon_1=\frac{P_T}{\sum_{k=1}^{K}\hat{p}_k}$
		and
		$\varepsilon_2=\frac{P_T}{\sum_{k=1}^{K}\hat{\beta}_k}$.
		\State {\textbf{for}} $k=1:K$, {\textbf{do}}
		\State ~~ Develop BB precoder vector ${\bf b}_k$ via \eqref{eq_BB_vector_2}.
		\State {\textbf{end for}}
		\Ensure ${\bf B}=\left[{\bf b}_1,\cdots,{\bf b}_K\right]$.
	\end{algorithmic}
\end{algorithm}

\section{Illustrative Results}\label{sec_Results}
The sum-rate, energy-efficiency, and fairness performance evaluation of the proposed NI-$\alpha$HP technique is presented throughout this section.
		Based on the recent 3GPP Release 17 \cite{Report_5G_Rel17_Mt_256,Report_5G_Macro_PL_Rel_17,Report_5G_UMi_UMa_Rel17}, the simulation setup for the MU-mMIMO systems is summarized in Table \ref{table1_sim}.
	On the other hand, Table \ref{table2_hyper_param} expresses the hyper-parameters for each proposed NI-$\alpha$HP technique, where GWO-$\alpha$HP does not have any hyper-parameters.
Also, NI-$\alpha$HP employs $N_S=100$ search agents through $Q=10$ iterations unless otherwise stated.

The sum-rate capacity in the MU-mMIMO systems is calculated as
$
	R_{\textrm{sum}} = \sum_{k=1}^{K}R_k
$ [bps/Hz].
Afterwards, the energy-efficiency is obtained by taking the ratio of the sum-rate capacity and the total transmit consumption as:
\begin{equation}
	\textrm{Energy-Efficiency} = \frac{R_{\textrm{sum}}}{P_T + N_{RF}\times P_{RF}} \textrm{ [bps/Hz/W]},
\end{equation}
where $P_{RF}=250$ mW denotes the power consumption per RF chain \cite{ANALOG_BF_Heath}. 
The number of RF chains utilized in the proposed NI-$\alpha$HP is provided in Table \ref{table1_sim}. On the other hand, the conventional single-stage FDP requires $N_{RF}=M=256$ RF chains, which deteriorates the energy-efficiency.

As a quantitative fairness measure, we here adopt Jain's fairness index\footnote{It is important to note that a larger value of Jain's fairness index indicates a higher fairness level (e.g., Jain's fairness index of $1$ implies absolute fairness).} based on each UE rate \cite{Fairness_Survey_2}:
\begin{equation}\label{eq_Jain}
	\textrm{Jain's Fairness Index}
	=
	\frac{\left(\sum_{k=1}^K R_k\right)^2}{K\sum_{k=1}^K R_k^2}\in\left[0,1\right].
\end{equation}

\begin{table}[t!]
	\caption{Simulation parameters.}
	\label{table1_sim}
	\centering
	\renewcommand{\arraystretch}{1.25}
	\begin{tabular}{|l|l|}
		\hline
		{\# of antennas \cite{Report_5G_Rel17_Mt_256}} & $M=16\times 16 = 256$      \\ \hline
		{BS transmit power  \cite{Report_5G_Macro_PL_Rel_17}} &  $P_T=20$ dBm or $P_T=40$ dBm \\ \hline
		{Cell radius  \cite{Report_5G_Rel17_Mt_256}} &  100m\\ \hline
		{BS height \big|\big. UE height \cite{Report_5G_UMi_UMa_Rel17}} &  10m \big|\big. 1.5m-2.5m \\ \hline
		{UE-BS horizontal distance} & 10m -- 100m\\ \hline
		{UE groups} & {$G=2$ or $G=4$}\\ \hline
		{UE per group} & $K_g=\frac{K}{G}$      \\ \hline
		{RF chains per group} & $N_{RF,g}=\frac{N_{RF}}{G}=8$      \\ \hline
		{Mean EAoD \& AAoD} & {$\theta_g \hspace{-0.5ex}=\hspace{-0.25ex} 50^\circ$,  $\psi_g\hspace{-0.5ex}=\hspace{-0.5ex}25^\circ 
			+
			\hspace{-0.5ex}
			\frac{360^\circ}{G}
			\left(g\hspace{-0.25ex}-\hspace{-0.25ex}1\right)
			$ }  \\ \hline
		EAoD \& AAoD spread \cite{Report_5G_UMi_UMa_Rel17}& 
		$\delta_g^\theta = \delta_g^\psi=10^\circ$ \\ \hline
		{Path loss exponent} \cite{Report_5G_Macro_PL_Rel_17}& $\eta=3.76$      \\ \hline
		{Noise PSD} \cite{Report_5G_Macro_PL_Rel_17}& $-174$ dBm/Hz      \\ \hline
		{Channel bandwidth} \cite{Report_5G_Macro_PL_Rel_17}& $120$ kHz      \\ \hline
		{\# of paths\cite{Report_5G_UMi_UMa_Rel17}} & $L=20$      \\ \hline
		{\# of network realizations} & $5000$      \\ \hline
	\end{tabular}
\end{table}

\begin{table}[t!]
	\caption{Hyper-parameters for NI-$\alpha$HP.}
	\label{table2_hyper_param}
	\centering
	\renewcommand{\arraystretch}{1.25}
	\begin{tabular}{|l|l|}
		\hline
		PSO-$\alpha$HP
		&
		$\kappa_{\Delta,\max}^{\textrm{PSO}}=-\kappa_{\Delta,\min}^{\textrm{PSO}}=-0.2$ \cite{ASIL_PSO_PA_WCNC}
		\\ \hline
		GWO-$\alpha$HP
		&
		-
		\\ \hline
		ACO-$\alpha$HP
		&
		$\kappa_1^{\textrm{ACO}}=10$,
		$\kappa_2^{\textrm{ACO}}=0.5$ \cite{ACO_cont} 
		\\ \hline
		CS-$\alpha$HP
		&
		$\kappa_1^{\textrm{CS}}=0.25$,
		$\kappa_2^{\textrm{CS}}=1.5$ \cite{PSO_book}
		\\ \hline
		FA-$\alpha$HP
		&
		$\kappa_1^{\textrm{FA}}=0.1$,
		$\kappa_2^{\textrm{FA}}=0.97$ \cite{PSO_book}
		\\ \hline
	\end{tabular}
\end{table}

\subsection{Benchmark on Sum-Rate Maximization}

\begin{figure}[!t]
	\centering
	\subfigure[Sum-rate]
	{\includegraphics[width=0.48\columnwidth]{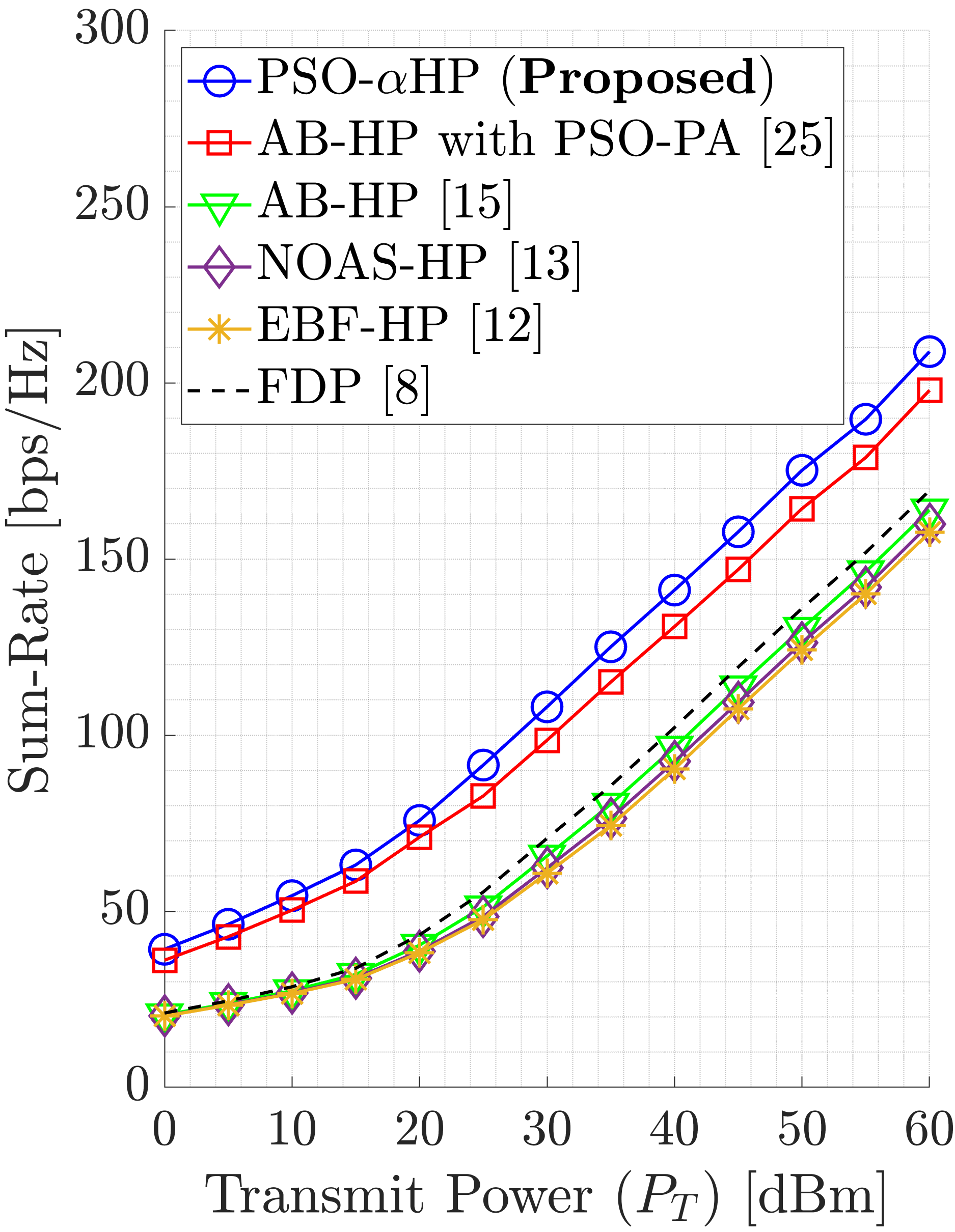}
		\label{fig_3a}}
	\subfigure[Energy-efficiency]
	{\includegraphics[width=0.48\columnwidth]{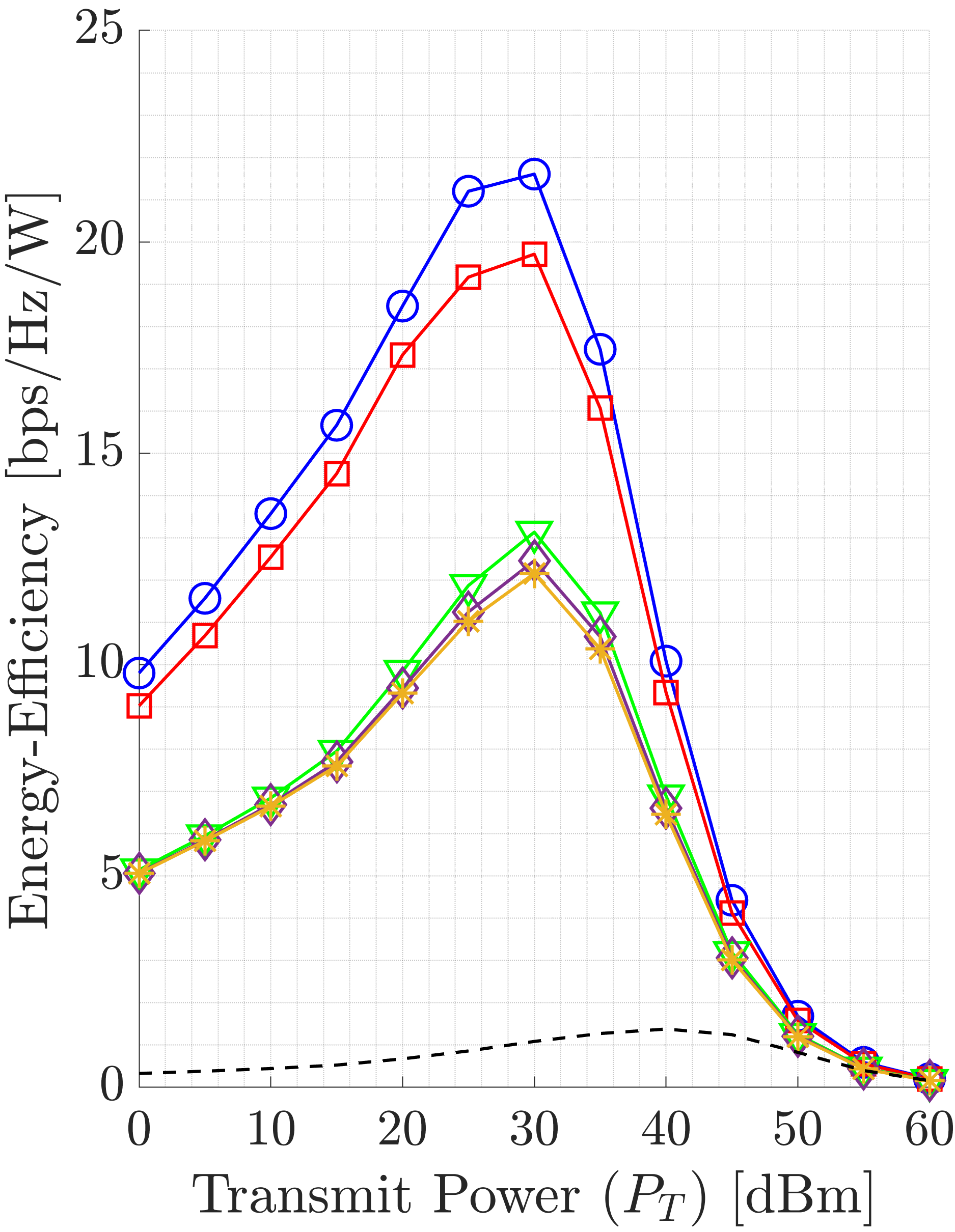}
		\label{fig_3b}}
	\caption{Sum-rate and energy-efficiency benchmark performance versus transmit power ($\alpha=0$, $G=2$ groups, $K=10$ UEs).
	}
	\label{fig_3}
\end{figure}

Fig. \ref{fig_3} plots the sum-rate and energy-efficiency performance of the proposed PSO-$\alpha$HP technique versus the transmit power, where 
	$K=10$ UEs are clustered in $G=2$ groups.
Here, we set $\alpha=0$ for sum-rate maximization objective as shown in \eqref{eq_alpha_fair} and \eqref{eq_obj_3_BB}.
	For the benchmark, we consider the conventional single-stage FDP \cite{MassMIMO_Precoding_2021} and various two-stage HP techniques such as 
		angular-based HP (AB-HP)\cite{ASIL_ABHP_Access},
		AB-HP with PSO-PA\footnote{AB-HP with PSO-PA
			develops the BB precoder by concatenating two sub-blocks: (i) the well-known RZF, (ii) PSO-PA optimizing $K$ power values. On the other hand, the proposed PSO-$\alpha$HP has a single-shot solution for the BB precoder given in \eqref{eq_BB_vector_2} by jointly optimizing $2K$ parameters.
		}\cite{ASIL_PSO_PA_WCNC},
		eigen-beamforming based HP (EBF-HP)\footnote{EBF-HP employs both phase-shifter and variable-gain amplifiers in the RF-stage, while the proposed NI-$\alpha$HP only uses the low-cost phase-shifters.} \cite{JSDM_LargeArray},
		non-orthogonal angle space based HP (NOAS-HP)\footnote{Unlike the proposed NI-$\alpha$HP, NOAS-HP needs the fast time-varying full CSI in the RF beamformer design \cite{MassMIMO_hybrid_NO_ADMA}. Nevertheless, it serves as a benchmark.} \cite{MassMIMO_hybrid_NO_ADMA}.
According to the simulation setup given in Table \ref{table1_sim}, FDP requires $M=256$ RF chains, whereas all HP schemes employ only $N_{RF} = 16$ RF chains.
	Hence, it indicates $93.75\%$ reduction in the hardware cost/complexity in the mMIMO systems.
The proposed NI-$\alpha$HP technique also lowers the channel estimation overhead size by $93.75\%$ (i.e., it employs $\bm{\mathcal{H}}\in\mathbb{C}^{10\times 16}$ instead of ${\bf{H}}\in\mathbb{C}^{10\times 256}$).

In Fig. \ref{fig_3a}, we first investigate the sum-rate performance,
		where all benchmark schemes, except AB-HP with PSO-PA, apply equal PA.
The numerical results reveal that the proposed PSO-$\alpha$HP greatly enhances the sum-rate capacity.
	To illustrate, when $P_T=30$ dBm, the sum-rate capacity for PSO-$\alpha$HP is 
		$108.9$ bps/Hz,	
	which is approximately 
	$10.4$ bps/Hz higher than AB-HP with PSO-PA,
	$39.5$ bps/Hz higher than FDP,
	$44.6$ bps/Hz higher than AB-HP,
	$48.7$ bps/Hz higher than NOAS-HP,
	$50.7$ bps/Hz higher than EBF-HP.
Moreover, the performance gap remains almost constant after $P_T=30$ dBm.
	On the other hand, it is seen that AB-HP outperforms both NOAS-HP and EBF-HP with the performance gap of $1.25$ dB and $1.75$ dB, respectively, in terms of the transmit power.
Furthermore, AB-HP only experiences $1.5$ dB degradation compared to the single-stage FDP.

	In Fig. \ref{fig_3b}, the energy-efficiency curves are presented versus the transmit power.
By means of a significant reduction in the number of RF chains, all HP schemes remarkably improve the energy-efficiency compared to the single-stage FDP.
	Furthermore, the highest energy-efficiency is attained via the proposed PSO-$\alpha$HP across all transmit power regimes.
For example, the energy-efficiency at $P_T=30$ dBm as $21.6$ bps/Hz/W for PSO-$\alpha$HP and $19.5$ bps/Hz/W for AB-HP with PSO-PA, however, it is only $1.1$ bps/Hz/W for FDP.

\begin{figure}[!t]
	\centering
	\includegraphics[width=\columnwidth]{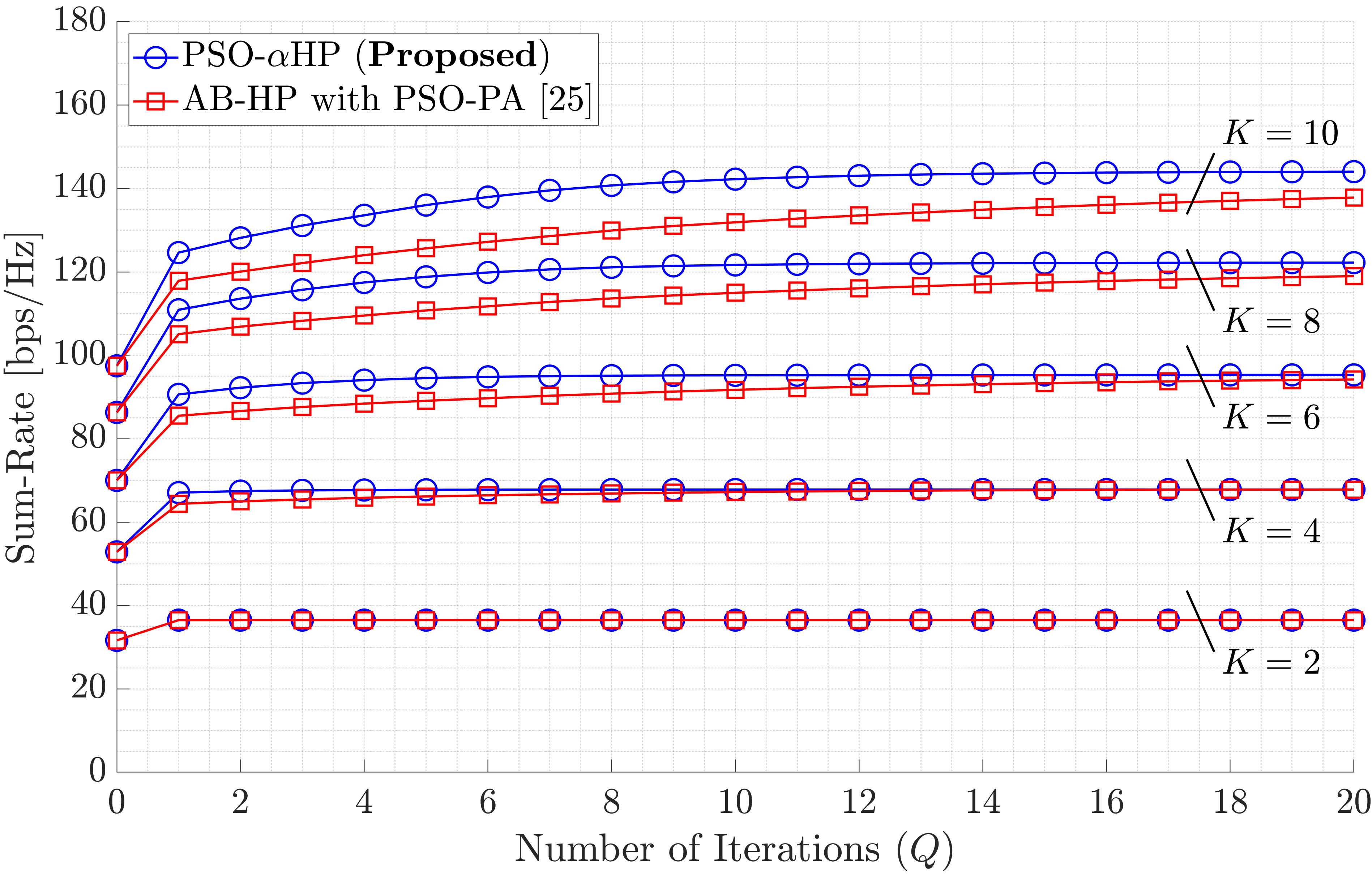}
	\caption{Sum-rate benchmark performance versus number of iterations ($\alpha=0$, $G=2$ groups, $P_T=40$ dBm).
	}
	\label{fig_4}
\end{figure}

Fig. \ref{fig_4} presents the sum-rate performance versus the number of iterations, where
	the transmit power is ${P_T=40}$ dBm to serve either $K=2,4,6,8$ or $10$ UEs clustered in $G=2$ groups.
	 For the sum-rate maximization objective, the fairness level  is kept as $\alpha=0$ to compare PSO-$\alpha$HP and AB-HP with PSO-PA. 
		Both techniques are investigated up to $Q=20$ iterations, while $Q=0$ indicates the AB-HP with equal PA.
The promising numerical results show that even though PSO-$\alpha$HP optimizes twice as many parameters as comparison to AB-HP with PSO-PA, it both converges faster and achieves higher capacity, especially for the larger number of UEs.
		For example, PSO-$\alpha$HP provides $3.5$ bps/Hz, $6.8$ bps/Hz or $10.4$ bps/Hz higher sum-rate at $Q=10$ iterations, when there are $K=6$, $8$ or $10$ UEs, respectively.
Moreover, the sum-rate performance of PSO-$\alpha$HP is approximately saturated within $Q=10$ iterations for $K=6,8,10$ UEs, whereas AB-HP with PSO-PA requires more than $Q=20$ iterations. 
	The main reason for this superior performance is the derived optimal digital BB precoder expression given in \eqref{eq_BB_vector}.
On the other hand, when there are only $K=2$ or $4$ UEs, both techniques converge in the early iterations.
	It is also important to highlight the sum-rate improvement accomplished by the proposed PSO-$\alpha$HP.  the AB-HP with equal PA (i.e., $Q=0$), PSO-$\alpha$HP improves the sum-rate capacity by
		$15.4\%$,
		$28.3\%$,
		$36.1\%$,
		$41.6\%$,
		$47.7\%$
		 for
		$K=2,4,6,8,10$ UEs, respectively.
Attractively, the sum-rate capacity improvement increases for the larger number of UEs.

\subsection{Performance Evaluation of NI-$\alpha$HP Techniques}
	
\begin{figure}[!t]
	\centering
	\includegraphics[width=\columnwidth]{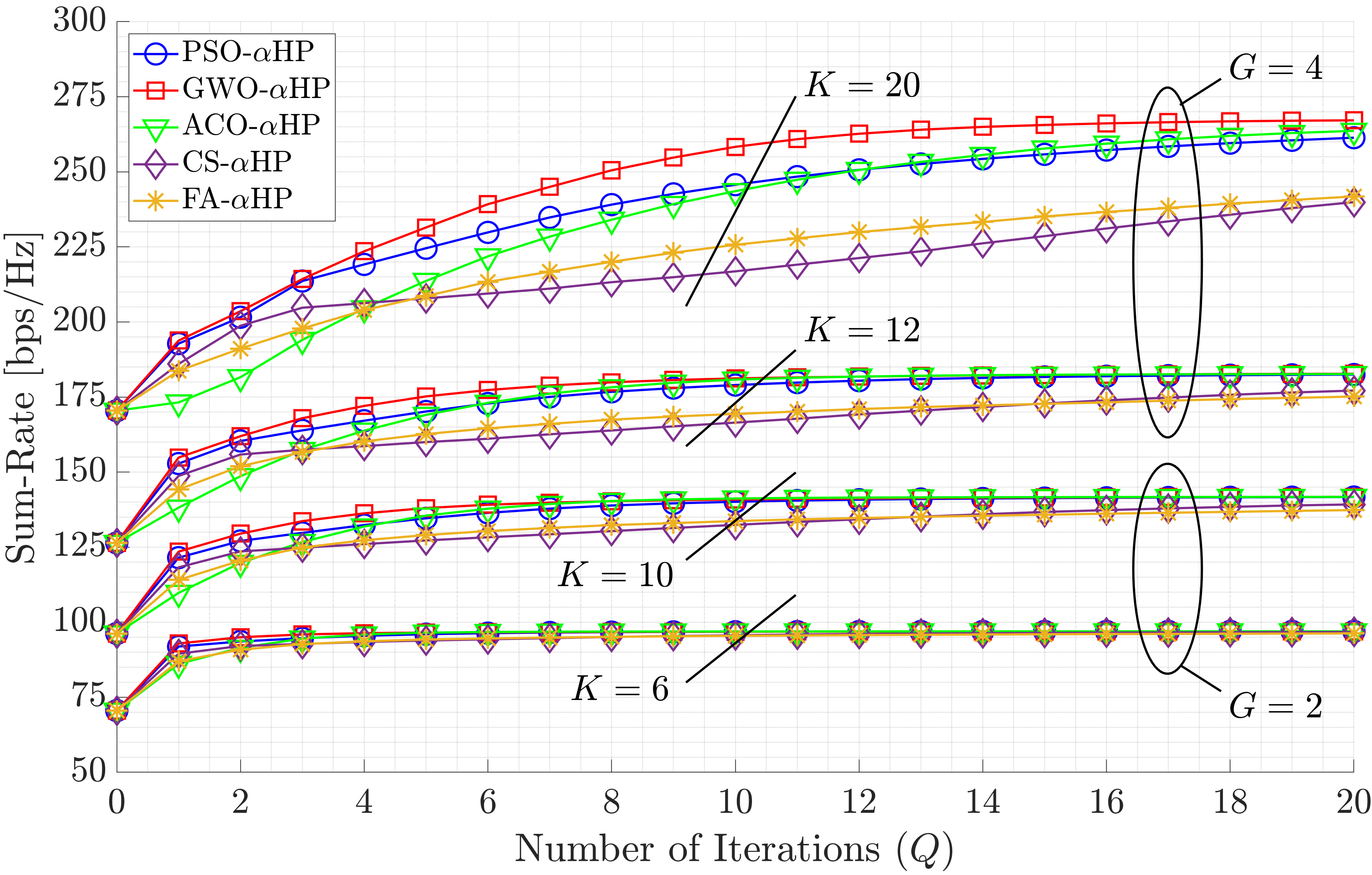}
	\caption{Sum-rate versus number of iterations ($P_T=40$ dBm and $\alpha=0$).}
	\label{fig_5}
\end{figure}

Fig. \ref{fig_5} compares all five NI-$\alpha$HP techniques in regards to their sum-rate performance versus the number of iterations, where $\alpha=0$ and $P_T=40$ dBm. 
Here, we serve either $K=6,10$ UEs in $G=2$ groups or $K=12,20$ UEs  in $G=4$ groups.
In every scenario, GWO-$\alpha$HP attains a higher sum-rate and converges faster than all
	PSO-$\alpha$HP,
	ACO-$\alpha$HP,
	CS-$\alpha$HP
		and
	FA-$\alpha$HP. 	
For instance, the sum-rate performance of GWO-$\alpha$HP is saturated around $Q=15$ iterations for $K=20$ UEs, while others require more than $Q=20$ iterations.	
	The intuitive explanation of these observations is that GWO-$\alpha$HP diversifies the search agents and increases exploration in the optimization space by asking them to simultaneously follow three best solutions (i.e., alpha, beta, and delta wolves), which also brings a beneficial competition.
On the other hand, each search agent in PSO-$\alpha$HP only follows the best solution achieved until the corresponding iteration, which might severely damage the competition and make them stuck around a local optima for long iterations.
	Similarly, each search agent in ACO-$\alpha$HP tracks only one of the randomly chosen $\kappa_1^{\textrm{ACO}}=10$ best solutions, which brings the competition among the best solutions and improves the performance of ACO-$\alpha$HP in the later iterations (e.g., ACO-$\alpha$HP outperforms PSO-$\alpha$HP, as $Q$ increases). However, compared to other four NI-$\alpha$HP techniques, ACO-$\alpha$HP experiences slow improvements in the early iterations due to following only one of the randomly chosen best solutions, which might be highly likely a local optimum.
		For example, ACO-$\alpha$HP achieves a lower sum-rate performance for $Q<6$ and $Q<12$ iterations compared to PSO-$\alpha$HP, when there are $K=12$ and $K=20$ UEs, respectively.
	On the other hand, we observe that both CS-$\alpha$HP and FA-$\alpha$HP experience comparatively low sum-rate improvements with respect to the other schemes.	
Additionally, by combining Fig. \ref{fig_4} and Fig. \ref{fig_5}, one can conclude that GWO-$\alpha$HP performs also better than AB-HP with PSO-PA \cite{ASIL_PSO_PA_WCNC}.

Table \ref{table3_runtime} displays the runtime performance evaluation\footnote{Each NI-$\alpha$HP scheme is implemented in MATLAB via a PC with Intel(R) Xeon(R) Gold 5220R CPU @ 2.20 GHz and 96 GB RAM.} for each NI-$\alpha$HP scheme. In addition to its superior performance, GWO-$\alpha$HP has the shortest runtime among all five schemes.
	On the contrary, CS-$\alpha$HP has the longest runtime in addition to its inadequate sum-rate performance presented in Fig. \ref{fig_5}.
Thus, GWO-$\alpha$HP is considered for the following  $\alpha$-fairness illustrative results based on its favorable performance.

\begin{table}[t!]
	\caption{Runtime per iteration.}
	\label{table3_runtime}
	\centering
	\renewcommand{\arraystretch}{1.25}
	\begin{tabular}{|l||rr|rr|}
		\hline
		\multirow{2}{*}{} & 
		\multicolumn{2}{c|}{$G=2$ groups} & 
		\multicolumn{2}{c|}{$G=4$ groups} 
		\\ \cline{2-5}
		&
		\multicolumn{1}{c|}{$6$ UEs} &
		\multicolumn{1}{c|}{$10$ UEs} &
		\multicolumn{1}{c|}{$12$ UEs} &
		\multicolumn{1}{c|}{$20$ UEs}
		\\ \hline \hline
		PSO-$\alpha$HP &
		\multicolumn{1}{r|}{4.0 msec} &
		\multicolumn{1}{r|}{4.4 msec} &
		\multicolumn{1}{r|}{6.4 msec} &
		\multicolumn{1}{r|}{7.6 msec}
		\\ \hline
		GWO-$\alpha$HP &
		\multicolumn{1}{r|}{\cellcolor{yellow} 3.5 msec} &
		\multicolumn{1}{r|}{\cellcolor{yellow} 3.9 msec} &
		\multicolumn{1}{r|}{\cellcolor{yellow} 6.0 msec} &
		\multicolumn{1}{r|}{\cellcolor{yellow} 7.3 msec} 
		\\ \hline
		ACO-$\alpha$HP&
		\multicolumn{1}{r|}{3.7 msec} &
		\multicolumn{1}{r|}{4.2 msec} &
		\multicolumn{1}{r|}{6.2 msec} &
		\multicolumn{1}{r|}{7.5 msec}
		\\ \hline
		CS-$\alpha$HP&
		\multicolumn{1}{r|}{7.8 msec}&
		\multicolumn{1}{r|}{8.3 msec}&
		\multicolumn{1}{r|}{11.7 msec}&
		\multicolumn{1}{r|}{14.5 msec}
		\\ \hline
		FA-$\alpha$HP&
		\multicolumn{1}{r|}{4.3 msec}&
		\multicolumn{1}{r|}{4.7 msec}&
		\multicolumn{1}{r|}{6.9 msec}&
		\multicolumn{1}{r|}{8.3 msec}
		\\ \hline
	\end{tabular}
\end{table}


\subsection{$\alpha$-Fairness}

\begin{figure*}[!t]
	\centering
	\subfigure[Sum-rate]
	{\includegraphics[width=\columnwidth]{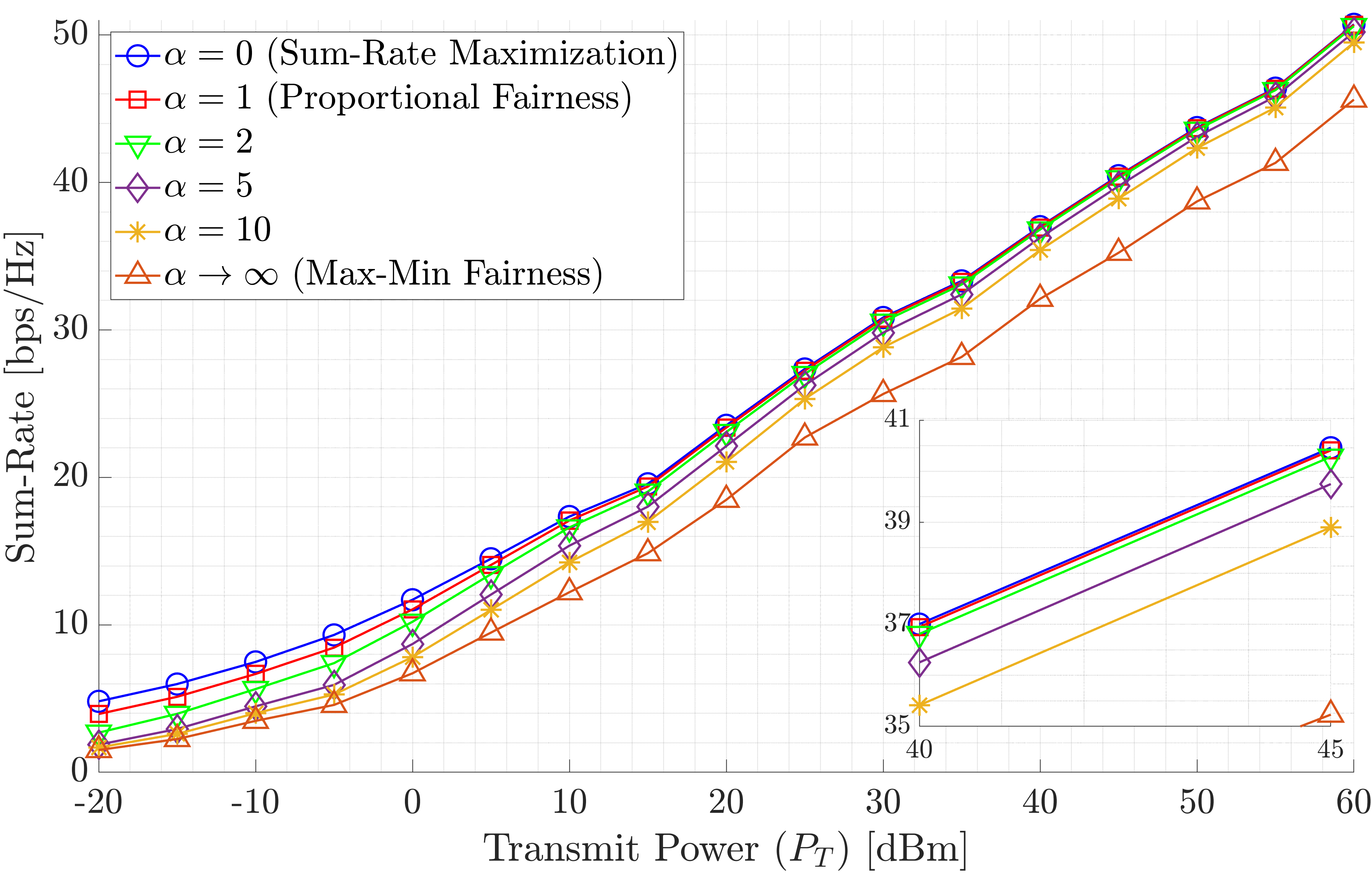}
		\label{fig_6a}}
	\subfigure[Jain's fairness index]
	{\includegraphics[width=0.48\columnwidth]{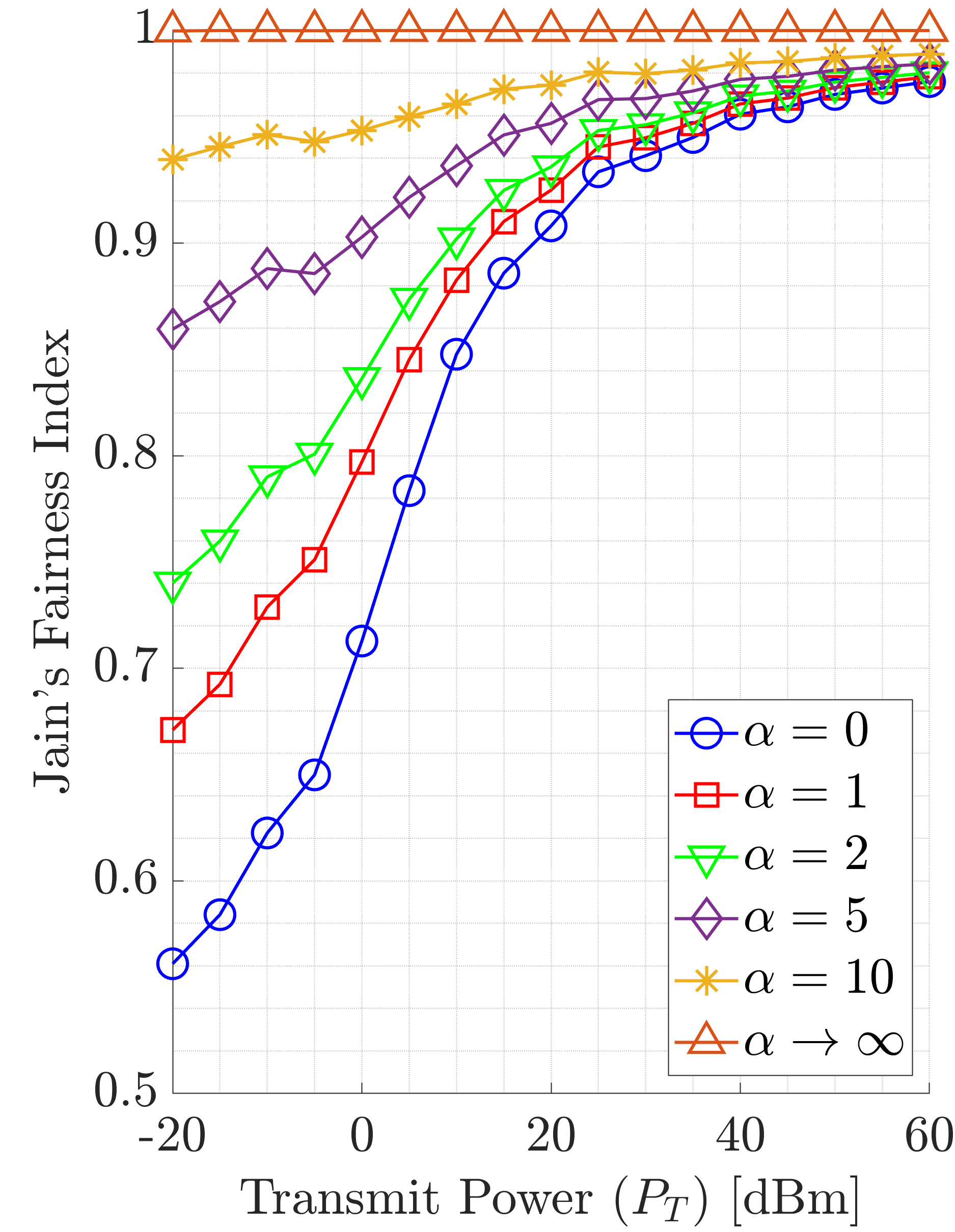}
		\label{fig_6b}}
	\subfigure[Rate gap]
	{\includegraphics[width=0.48\columnwidth]{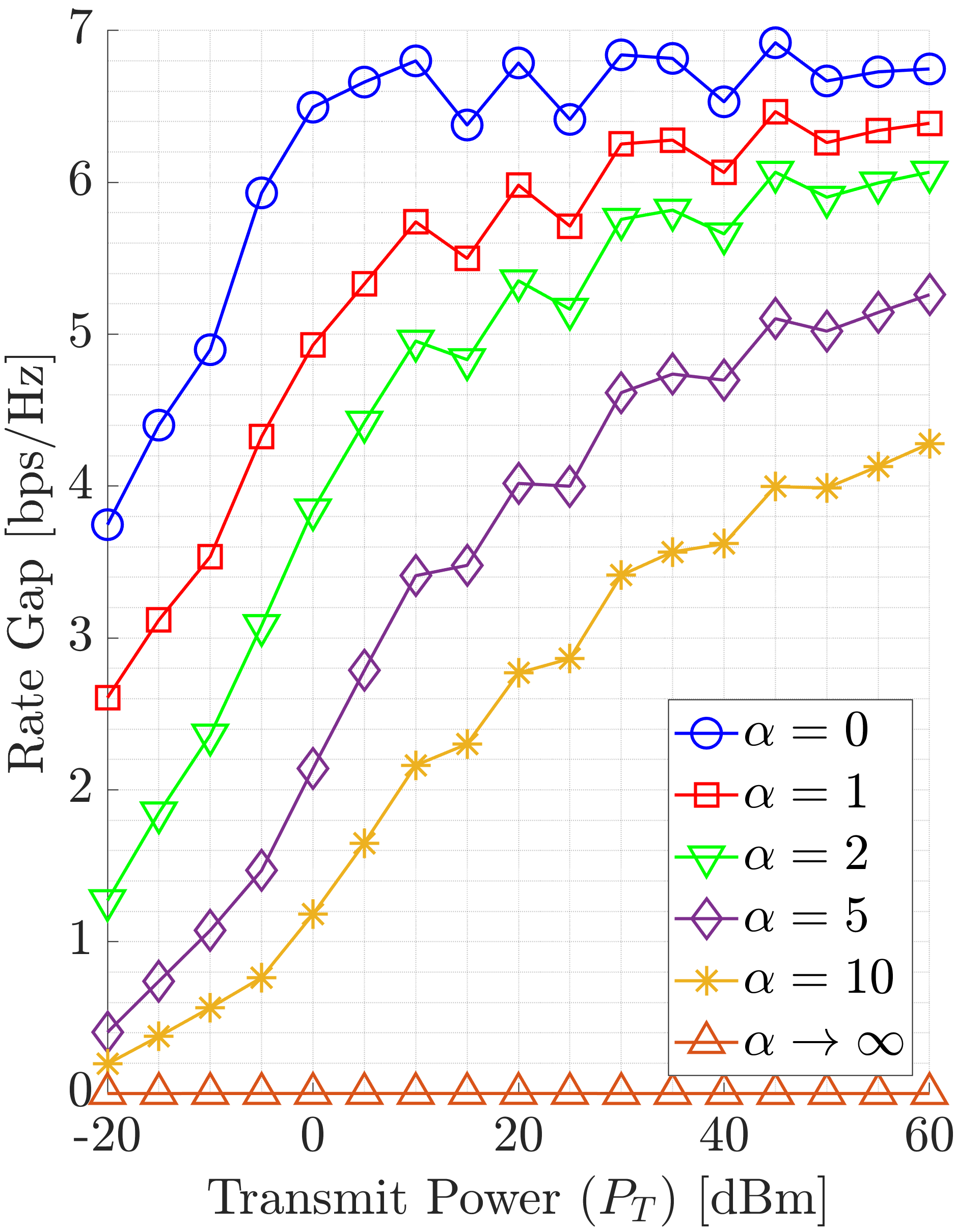}
		\label{fig_6c}}
	\caption{Sum-rate, fairness, and rate-gap performance of GWO-$\alpha$HP versus transmit power ($G=2$ groups, $K=2$ UEs).
	}
	\label{fig_6}
\end{figure*}

\begin{figure*}[!t]
	\centering
	\subfigure[Jain's fairness index]
	{\includegraphics[width=\columnwidth]{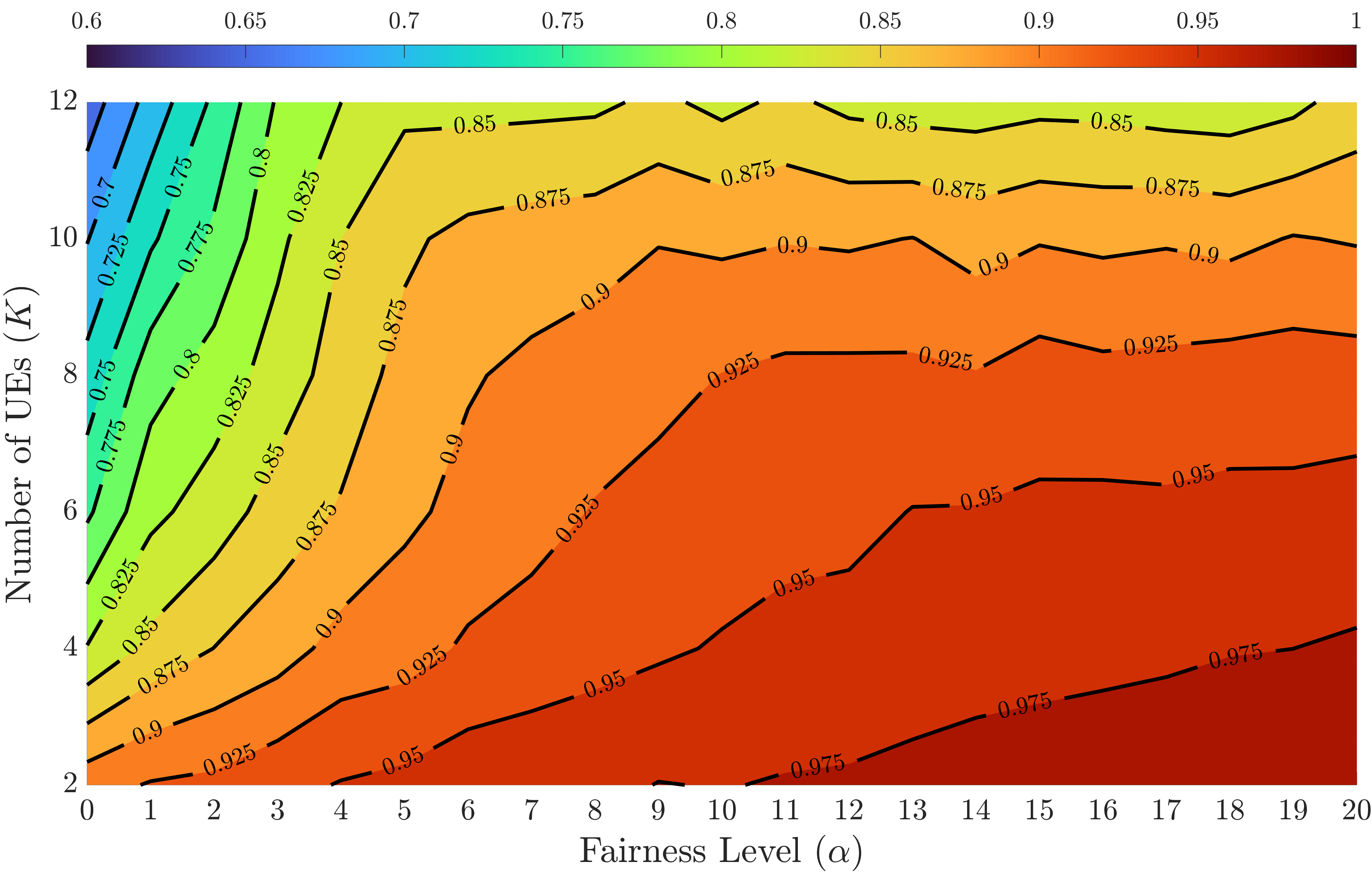}
		\label{fig_7a}}
	\subfigure[Rate gap]
	{\includegraphics[width=\columnwidth]{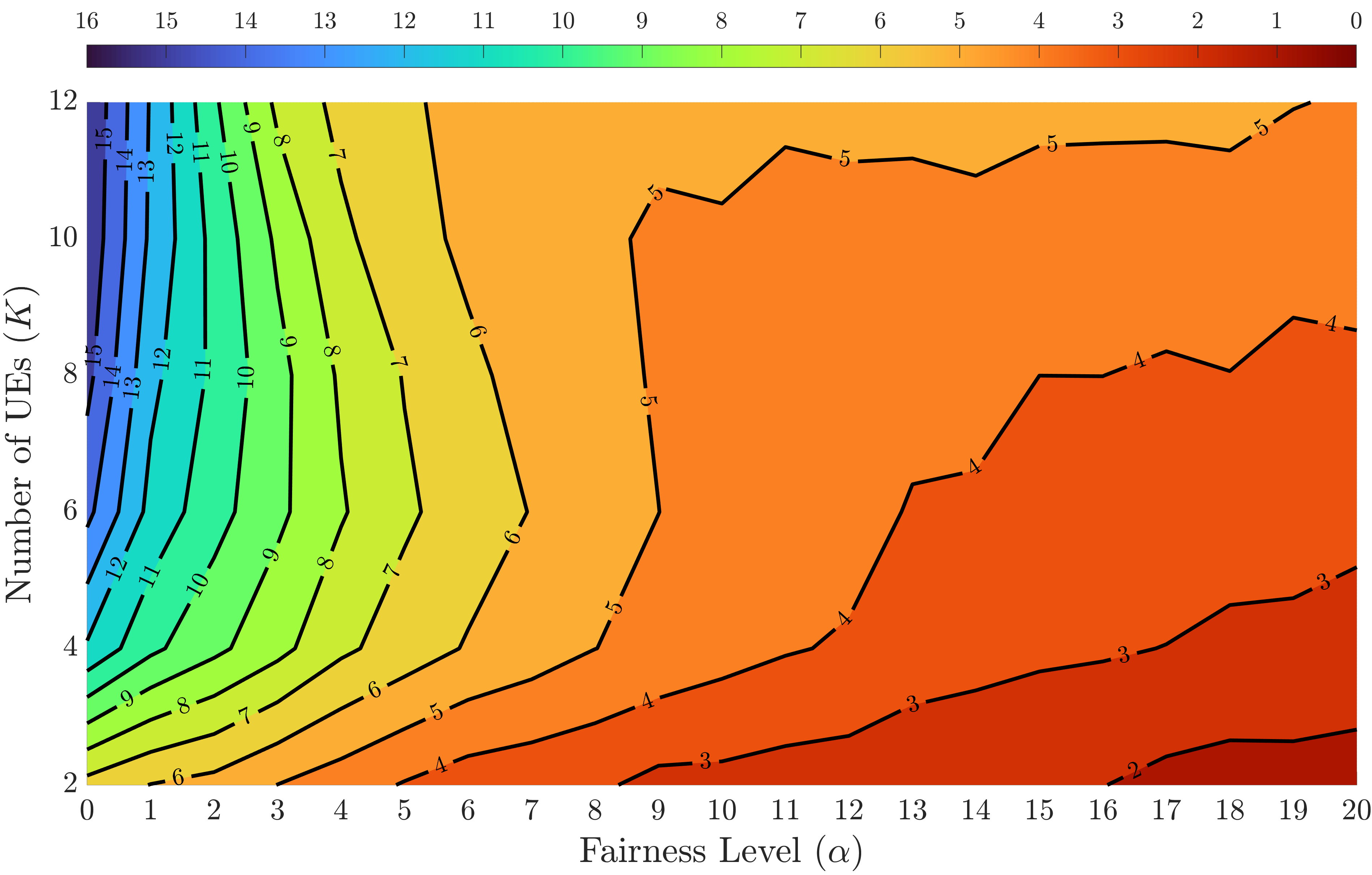}
		\label{fig_7b}}
	\caption{Fairness and rate-gap performance of GWO-$\alpha$HP versus number of UEs and fairness level ($G=2$ groups, $P_T=20$ dBm).
	}
	\label{fig_7}
	\vspace{-1ex}
\end{figure*}

Fig. \ref{fig_6} investigates the proposed GWO-$\alpha$HP technique for various fairness levels such as $\alpha=0,1,2,5,10,\infty$, where there are only $K=2$ UEs in $G=2$ groups.
As expressed in Section \ref{sec_ProblemForm}, $\alpha=0$ is for the sum-rate maximization, ${\alpha=1}$ indicates the proportional fairness, while ${\alpha\to\infty}$ implies the max-min fairness.
	Specifically, we illustrate the sum-rate and Jain's fairness index performance in Fig. \ref{fig_6a} and Fig. \ref{fig_6b}, respectively.
By defining ${R_{\max} =\max_k R_k}$ and ${R_{\min} = \min_k R_k}$, Fig. \ref{fig_6c} also plots the rate gap given by:
\begin{equation}
	\textrm{Rate Gap} = R_{\max}-R_{\min}  \textrm{ [bps/Hz]}.
\end{equation}
In Fig. \ref{fig_6a}, we observe a sum-rate performance degradation for the higher values of $\alpha$ as a result of increased fairness level. 
For example, the sum-rate at $P_T=0$ dBm drops from
	$11.7$ bps/Hz to $6.7$ bps/Hz, 
when the fairness level is increased from $\alpha=0$ to $\alpha\to \infty$.
On the other hand, as $\alpha$ increases, Jain's fairness index given in \eqref{eq_Jain} improves as demonstrated in Fig. \ref{fig_6b}.
For instance, the fairness index at $P_T=0$ dBm is improved from $0.71$ to $0.81$ by switching from the sum-rate maximization to the proportional fairness,
	while the max-min fairness 
	achieves the fairness index of $1.0$.
The numerical results also reveal that the proposed GWO-$\alpha$HP technique can address various fairness expectations by adjusting $\alpha$.
		It is seen that GWO-$\alpha$HP even supports absolute fairness (i.e., Jain's fairness index of $1.0$).
	Furthermore, although the sum-rate capacity linearly increases in the high transmit power regime as seen in Fig. \ref{fig_6a}, the rate gap among $K=2$ UEs is almost saturated as shown in Fig. \ref{fig_6c}. Hence, when the transmit power increases, Jain's fairness index also improves as presented in Fig. \ref{fig_6b}. 
		Additionally, the rate gap decays for the higher fairness levels and it even vanishes for $\alpha\to\infty$.

Fig. \ref{fig_7} illustrates Jain's fairness index and rate gap contour plots versus the number of UEs and the fairness level, where we set the transmit power $P_T=20$ dBm to support either $K=2,4,6,8,10$ or $12$ UEs in $G=2$ groups.
	Also, the fairness levels are selected as $\alpha=0,1,2,\cdots,20$.
When GWO-$\alpha$HP targets the sum-rate maximization objective function (i.e., $\alpha=0$), the rate gap remarkably increases for the larger number of UEs, which also deteriorates Jain's fairness index.
	To illustrate, when the number of UEs is changed from $K=2$ to $K=12$, the rate gap enlarges from $6.6$ bps/Hz to $15.9$ bps/Hz and the fairness index sharply reduces from $0.91$ to $0.66$, respectively.
On the other hand, GWO-$\alpha$HP jointly reduces the rate gap and improves Jain's fairness index by selecting a higher fairness level.
	For example, when the fairness level is set to $\alpha=20$ for $K=12$ UEs, the rate gap is calculated as only $4.8$ bps/Hz and Jain's fairness index is enhanced to $0.86$.
As mentioned earlier, it is possible to adjust $\alpha$ based on the fairness index expectation. To illustrate, when the desired Jain's fairness index is at least $0.9$ for $K=8$ UEs, it is necessary to choose $\alpha\ge 7$.

\section{Conclusions}\label{sec_Conc}
In this work, a novel nature-inspired $\alpha$-fair hybrid precoding (NI-$\alpha$HP) technique has been proposed for the mmWave MU-mMIMO systems.
	First,	the analog RF beamformer has been developed via slow time-varying AoD information for both reducing the channel estimation overhead size and  maximizing the beamforming gain in the desired direction.
Second, the effective channel seen from the BB-stage has been employed to design the digital BB precoder. 
	Based on the $\alpha$-fair resource allocation objective, we have derived the optimal digital BB precoder expression, which includes a set of {NP-hard} parameters.
Afterwards, they have been effectively optimized via applying five nature-inspired intelligent algorithms. Particularly, we have introduced PSO-$\alpha$HP, GWO-$\alpha$HP, ACO-$\alpha$HP, CS-$\alpha$HP, FA-$\alpha$HP.
	The numerical results reveal that under the sum-rate maximization objective, the proposed NI-$\alpha$HP techniques outperform other benchmark HP schemes in terms of both sum-rate and energy-efficiency performance.
Furthermore, we have observed that GWO-$\alpha$HP achieves a higher sum-rate and converges faster among the five proposed NI-$\alpha$HP techniques.
	Regarding the fairness perspective, NI-$\alpha$HP successfully reduces the rate gap among UEs and satisfies various fairness expectations by simply adjusting the fairness level ($\alpha$).
	
\appendices

\section{Proof of Proposition 1}\label{app_NOBF_A}
By using the $\alpha$-fairness utility function given in \eqref{eq_alpha_fair}, we first write the Lagrangian function of \eqref{eq_obj_2_BB} for $\alpha\ne 1$ as:
\begin{equation}\label{eq_Lagrangian}
\begin{aligned}
	\mathcal{L}&
		\left(
			{\bf b}_1,
			\cdots,
			{\bf b}_K,
			\bar{\beta}_0,
			\bar{\beta}_1,
			\cdots,
			\bar{\beta}_K
		\right)\\
	&=
	\sum_{k=1}^{K}
	\log_2
	\left(
		1+
		\frac
		{\left|{\bf h}_k^T{\bf Fb}_k\right|^2}
		{\sum_{u\ne k}^{K}\left|{\bf h}_k^T{\bf Fb}_u\right|^2 
			+
		\sigma_n^2}
	\right)^{1-\alpha}\\
	&+
	\bar{\beta}_0
	\left(
		\sum_{k=1}^{K}{\bf b}_k^H {\bf b}_k - P_T
	\right)\\
	&+
	\sum_{k=1}^{K}
	\bar{\beta}_k
	\left(
		1
		-
		\frac
		{\left|{\bf h}_k^T{\bf Fb}_k\right|^2}
		{\mu_k\sigma_n^2}
		+
		\frac
		{\sum_{u\ne k}^{K}\left|{\bf h}_k^T{\bf Fb}_u\right|^2}
		{\sigma_n^2}
	\right),
\end{aligned}
\end{equation}
where $\bar{\beta}_0\ge0$ and $\bar{\beta}_k\ge0$ are the Lagrangian multipliers associated with the transmit power constraint (i.e., $C_3$ in \eqref{eq_obj_2_BB}) and the $k^{th}$ UE SINR constraint (i.e., $C_2$ in \eqref{eq_obj_2_BB}), respectively. By applying the strong duality property \cite{strongDuality_Lagrangian}, the dual function is obtained as follows:
\begin{equation}
\begin{aligned}
	\min_{{\bf b}_1,\cdots,{\bf b}_K}
	\mathcal{L}&
	\left(
		{\bf b}_1,
		\cdots,
		{\bf b}_K,
		\bar{\beta}_0,
		\bar{\beta}_1,
		\cdots,
		\bar{\beta}_K
	\right)
	\\
	&
	=
	-
	\bar{\beta}_0 P_T
	+
	\sum_{k=1}^{K}
	\bar{\beta}_k
	=
	0,
\end{aligned}
\end{equation}
which implies $\sum_{k=1}^{K}\beta_k=P_T$ with $\beta_k=\frac{\bar{\beta}_k}{\bar{\beta}_0}$. 
	In order to find the optimal ${\bf b}_k$, we here exploit the KKT conditions. Thus, the gradient of the Lagrangian with respect to ${\bf b}_k$ is set to zero:
\begin{equation}\label{eq_Lagrangian_der}
	\begin{aligned}
		\frac{\partial
		\mathcal{L}
		\left(
		{\bf b}_1,
		\cdots,
		{\bf b}_K,
		\bar{\beta}_0,
		\bar{\beta}_1,
		\cdots,
		\bar{\beta}_K
		\right)
		}{\partial{\bf b}_k}
		=0,
		~
		\forall k.
	\end{aligned}
\end{equation}
By using \eqref{eq_SINR} and \eqref{eq_Lagrangian}, the Lagrangian function is expanded as
$\mathcal{L}
\left(
{\bf b}_1,
\cdots,
{\bf b}_K,
\bar{\beta}_0,
\bar{\beta}_1,
\cdots,
\bar{\beta}_K
\right)
=
\mathcal{L}_1
+
\mathcal{L}_2
+
\mathcal{L}_3
$ 
with
$\mathcal{L}_1
=
\sum_{u=1}^{K}
\log_2
\left(
	1
	+
	\textrm{SINR}_u
\right)^{1-\alpha}$,
$\mathcal{L}_2
=
\bar{\beta}_0
\left(
\sum_{u=1}^{K}{\bf b}_u^H {\bf b}_u - P_T
\right)
$ 
and
$\mathcal{L}_3
=
\sum_{u=1}^{K}
\bar{\beta}_u
\left(
1
-
\frac
{\left|{\bf h}_u^T{\bf Fb}_u\right|^2}
{\mu_u\sigma_n^2}
+
\frac
{\sum_{i\ne u}^{K}\left|{\bf h}_u^T{\bf Fb}_i\right|^2}
{\sigma_n^2}
\right)
$.
After some mathematical derivations, we find the gradient of $\mathcal{L}_1$ as:
\begin{equation}\label{eq_Lagrangian_1}
\begin{aligned}
	\frac{\partial \mathcal{L}_1}{\partial {\bf b}_k}
	&=
	\frac{\partial}{\partial {\bf b}_k}
	\hspace{-0.5ex}
	\left[
	\sum_{u=1}^{K}
	\hspace{-0.25ex}
	\log_2
	\hspace{-0.75ex}
	\left(
	\hspace{-0.5ex}
	1
	\hspace{-0.25ex}+\hspace{-0.25ex}
	\frac
	{\left|{\bf h}_u^T{\bf Fb}_u\right|^2}
	{\sum_{i\ne u}^{K}
		\hspace{-0.25ex}
		\left|{\bf h}_u^T{\bf Fb}_i\right|^2 
		\hspace{-0.5ex}+\hspace{-0.5ex}
		\sigma_n^2}
	\right)^{\hspace{-1ex}1-\alpha}
	\hspace{-0.25ex}
	\right]
	\\
	&
	=
	\frac{\partial}{\partial {\bf b}_k}
	\bigg[
		\log_2
		\left(
		1
		+
		\textrm{SINR}_k
		\right)^{1-\alpha}
		\bigg.\\
	&
	\bigg.+
		\sum_{u\ne k}^{K}
		\log_2
		\left(
		1
		+
		\textrm{SINR}_u
		\right)^{1-\alpha}
	\bigg]
	\\
	&=
	\bigg[
		\frac{2}{\chi _k} \log _2 \left( 1 + \textrm{SINR}_k \right)^{ - \alpha }
	\Big. \\
	&
	\bigg.
	+
		\sum_{u\ne k}^{K}
		\frac{2\times \textrm{SINR}_u}{\chi _u}
		{{\log }_2}{{\left( {1 + \textrm{SINR}_u} \right)}^{ - \alpha }}
	\bigg]
	{\bf A}_k {\bf b}_k,
\end{aligned}
\end{equation}
where 
	${\bf{A}}_k = {\bf F}^H {\bf h}_k^*{\bf h}_k^T{\bf F}$
and
	$\chi _k 
	= \frac
	{1 - \alpha }
	{\ln \left( 2 \right)\left( {\left\| {{\bf h}_k^T\bf{FB}} \right\|^2 + \sigma_n^2} \right)}$.
Afterwards,
	the gradient of $\mathcal{L}_2$ is obtained as follows:
\begin{equation}\label{eq_Lagrangian_2}
	\frac{\partial \mathcal{L}_2}{\partial {\bf b}_k}
	=
	\frac{\partial }{{\partial {\bf{b}}_k}}\left[ {\bar{\beta}_0\left( {\sum\limits_{u = 1}^K {{\bf{b}}_u^H{{\bf{b}}_u} - {P_T}} } \right)} \right] = 2\bar{\beta}_0{{\bf{b}}_k}.
\end{equation}
Moreover, the gradient of $\mathcal{L}_3$ is derived as:
\begin{equation}\label{eq_Lagrangian_3}
	\begin{aligned}
		&\frac{\partial \mathcal{L}_3}{\partial {\bf b}_k}
		\hspace{-.25ex}=\hspace{-.25ex}
		\frac{\partial}{\partial {\bf b}_k}
		\hspace{-.5ex}
		\left[
		\hspace{-.2ex}
		\sum_{u=1}^{K}
		\hspace{-.25ex}
		\bar{\beta}_u
		\hspace{-.75ex}
		\left(
		\hspace{-.5ex}
		1
		\hspace{-.5ex}-\hspace{-.5ex}
		\frac
		{\left|{\bf h}_u^T{\bf Fb}_u\right|^{\hspace{-0.2ex}2}\hspace{-0.75ex}}
		{\mu_u\sigma_n^2}
		\hspace{-.5ex}+\hspace{-.5ex}
		\frac
		{\sum_{i\ne u}^{K}
			\hspace{-.5ex}
			\left|{\bf h}_u^T{\bf Fb}_i\right|^{\hspace{-0.2ex}2}\hspace{-0.75ex}}
		{\sigma_n^2}
		\hspace{.5ex}
		\right)
		\hspace{-.5ex}
		\right]
		\\
	&
	=
	\frac{\partial}{\partial {\bf b}_k}
	\Bigg[
		\bar{\beta}_k
		\hspace{-.75ex}
		\left(
		\hspace{-.5ex}
		1
		\hspace{-.5ex}-\hspace{-.5ex}
		\frac
		{\left|{\bf h}_k^T{\bf Fb}_k\right|^{\hspace{-0.2ex}2}\hspace{-0.75ex}}
		{\mu_k\sigma_n^2}
		\hspace{-.5ex}+\hspace{-.5ex}
		\frac
		{\sum_{i\ne k}^{K}
			\hspace{-.5ex}
			\left|{\bf h}_k^T{\bf Fb}_i\right|^{\hspace{-0.2ex}2}\hspace{-0.75ex}}
		{\sigma_n^2}
		\hspace{.5ex}
		\right)
	\Bigg.\\
	&
	\Bigg.
	+
	\hspace{-0.75ex}
	\sum_{u\ne k}^{K}
	\hspace{-.5ex}
	\bar{\beta}_u
	\hspace{-.75ex}
	\left(
	\hspace{-.75ex}
	1
	\hspace{-.5ex}-\hspace{-.5ex}
	\frac
	{\left|{\bf h}_u^{\hspace{-0.1ex}T}\hspace{-.15ex}{\bf Fb}_u\right|^{\hspace{-0.2ex}2}\hspace{-1ex}}
	{\mu_u\sigma_n^2}
	\hspace{-.5ex}+\hspace{-.5ex}
	\frac
	{\left|{\bf h}_u^{\hspace{-0.1ex}T}\hspace{-.15ex}{\bf Fb}_k\right|^{\hspace{-0.2ex}2}\hspace{-1ex}}
	{\sigma_n^2}
	\hspace{-.5ex}+\hspace{-.5ex}
	\frac
	{\sum_{i\ne u,k}^{K}
		\hspace{-.5ex}
		\left|{\bf h}_u^{\hspace{-0.1ex}T}\hspace{-.15ex}{\bf Fb}_i\right|^{\hspace{-0.2ex}2}\hspace{-0.75ex}}
	{\sigma_n^2}
	\hspace{.5ex}
	\right)
	\hspace{-1.25ex}
	\Bigg]
	\\
	&
	=
	-
	\frac{2\bar{\beta}_k}{\mu_k\sigma_n^2}
	{\bf A}_k {\bf b}_k
	+
	\sum_{u\ne k}^{K}
	\frac{2\bar{\beta}_u}{\sigma_n^2}
	{\bf A}_u {\bf b}_k\\
	&
	=
	-
	\frac{2\bar{\beta}_k}{\sigma_n^2}
	\left(1+\frac{1}{\mu_k}\right)
	{\bf A}_k {\bf b}_k
	+
	\sum_{u=1}^{K}
	\frac{2\bar{\beta}_u}{\sigma_n^2}
	{\bf A}_u {\bf b}_k.
	\end{aligned}
\end{equation}
By substituting \eqref{eq_Lagrangian}, \eqref{eq_Lagrangian_1}, \eqref{eq_Lagrangian_2}, and \eqref{eq_Lagrangian_3} into \eqref{eq_Lagrangian_der},
the following expression can be written:
\begin{equation}\label{eq_Lagrangian_der_2}
	\begin{aligned}
		&\frac{\partial
			\mathcal{L}
			\left(
			{\bf b}_1,
			\cdots,
			{\bf b}_K,
			\bar{\beta}_0,
			\bar{\beta}_1,
			\cdots,
			\bar{\beta}_K
			\right)
		}{\partial{\bf b}_k}\\
		&
		=
		\bigg[
		-
		\frac{{\beta}_k}{\sigma_n^2}
		+
		\frac{1}{\chi _k\bar{\beta}_0} \log _2 \left( 1 + \textrm{SINR}_k \right)^{ - \alpha }
		\left(1+\frac{1}{\mu_k}\right)\\
		&+
		\sum_{u\ne k}^{K}
		\frac{\textrm{SINR}_u}{\chi _u\bar{\beta}_0}
		{{\log }_2}{{\left( {1 + \textrm{SINR}_u} \right)}^{ - \alpha }}
		\bigg]
		{\bf h}_k^T{\bf{F}}{\bf b}_k {\bf F}^H {\bf h}_k^*\\
		&
		+
		\bigg[
			{\bf I}_{N_{RF}}
			+
			\sum_{u=1}^{K}
			\frac{{\beta}_u}{\sigma_n^2}
			{\bf A}_u
		\bigg] 
		{\bf b}_k\\
		&=0
	\end{aligned}
\end{equation}
where we utilize 
$\frac{\partial \left(\mathcal{L}_1+\mathcal{L}_2+\mathcal{L}_3\right)}{\partial {\bf b}_k}
=
\left({2\bar{\beta}_0}\right)^{-1}
\frac{\partial \left(\mathcal{L}_1+\mathcal{L}_2+\mathcal{L}_3\right)}
{\partial {\bf b}_k}
=
0
$
and
$
{\bf A}_k {\bf b}_k
=
{\bf F}^H {\bf h}_k^*{\bf h}_k^T{\bf{F}}{\bf b}_k
=
{\bf h}_k^T{\bf{F}}{\bf b}_k {\bf F}^H {\bf h}_k^*
$.
Afterwards, we derive the optimal digital BB precoder vector for the $k^{th}$ UE as follows:
\begin{equation}\label{eq_APP_BB_vect}
	\begin{aligned}
		{\bf b}_k
		&
		=
		\bigg[
		\frac{{\beta}_k}{\sigma_n^2}
		-
		\frac{1}{\chi _k\bar{\beta}_0} \log _2 \left( 1 + \textrm{SINR}_k \right)^{ - \alpha }
		\left(1+\frac{1}{\mu_k}\right)\\
		&-
		\sum_{u\ne k}^{K}
		\frac{\textrm{SINR}_u}{\chi _u\bar{\beta}_0}
		{{\log }_2}{{\left( {1 + \textrm{SINR}_u} \right)}^{ - \alpha }}
		\bigg]
		{\bf h}_k^T{\bf{F}}{\bf b}_k\\
		&\times
		\bigg[
		{\bf I}_{N_{RF}}
		+
		\sum_{u=1}^{K}
		\frac{{\beta}_u}{\sigma_n^2}
		{\bf A}_u
		\bigg]^{-1}
		{\bf F}^H {\bf h}_k^*.\\
	\end{aligned}
\end{equation}
Then, one can define the allocated power for the $k^{th}$ UE as:
\begin{equation}\label{eq_APP_pk}
	\begin{aligned}
		{p_k}
		&={\bf b}_k^H{\bf b}_k\\
		&=
		{
			\bigg\|
			\bigg[
			{\bf I}_{N_{RF}} + \sum_{u=1}^K \frac{\beta_u}{\sigma^2_n} {\bf A}_u
			\bigg]^{-1}
			{\bf F}^H {\bf h}_k^*
			\bigg\|_2^2
		}\\
		&
		\times
		\bigg[
		\frac{{\beta}_k}{\sigma_n^2}
		-
		\frac{1}{\chi _k\bar{\beta}_0} \log _2 \left( 1 + \textrm{SINR}_k \right)^{ - \alpha }
		\left(1+\frac{1}{\mu_k}\right)\\
		&
		-
		\sum_{u\ne k}^{K}
		\frac{\textrm{SINR}_u}{\chi _u\bar{\beta}_0}
		{{\log }_2}{{\left( {1 + \textrm{SINR}_u} \right)}^{ - \alpha }}
		\bigg]^2
		\left({\bf h}_k^T{\bf{F}}{\bf b}_k\right)^2.
	\end{aligned}
\end{equation}

Finally, by combining \eqref{eq_APP_BB_vect} and \eqref{eq_APP_pk}, the optimal digital BB precoder vector is rewritten as follows:
\begin{equation}\label{eq_APP_BB_vect_2}
	\begin{aligned}
		{\bf b}_k
		&=
		\underbrace{\sqrt{p_k}}_{
			{\textrm{Power}}
		}
		\underbrace{
			\frac{
				\left[
				{\bf I}_{N_{RF}} + \sum_{u=1}^K \frac{\beta_u}{\sigma^2_n} {\bf A}_u
				\right]^{-1}
				{\bf F}^H {\bf h}_k^*
			}{
				\left\|
				\left[
				{\bf I}_{N_{RF}} + \sum_{u=1}^K \frac{\beta_u}{\sigma^2_n} {\bf A}_u
				\right]^{-1}
				{\bf F}^H {\bf h}_k^*
				\right\|
			}
		}_{
			\textrm{Beamforming Direction}
		},
	\end{aligned}
\end{equation}
where 
	$\sum_{k=1}^K p_k = P_T$
and 
	$\sum_{k=1}^K \beta_k = P_T$.
It is important to remark that we find the closed-from expression for the optimal digital BB precoder as shown in \eqref{eq_APP_BB_vect_2}, however, {finding the optimal values for $p_k$ and $\beta_k$ is NP-hard due to the entangled optimization parameters as expressed in \eqref{eq_APP_pk} \cite{Emil_PA_NonConvex}}.
	Moreover, one can easily show that \eqref{eq_APP_BB_vect_2} is valid for all fairness levels including $\alpha=1$.
Hence, this concludes the proof of \eqref{eq_BB_vector}.

	\ifCLASSOPTIONcaptionsoff
	\newpage
	\fi

	\bibliographystyle{IEEEtran}
	\bibliography{bibAsil_2205}

\begin{thebibliography}{10}
\providecommand{\url}[1]{#1}
\csname url@samestyle\endcsname
\providecommand{\newblock}{\relax}
\providecommand{\bibinfo}[2]{#2}
\providecommand{\BIBentrySTDinterwordspacing}{\spaceskip=0pt\relax}
\providecommand{\BIBentryALTinterwordstretchfactor}{4}
\providecommand{\BIBentryALTinterwordspacing}{\spaceskip=\fontdimen2\font plus
\BIBentryALTinterwordstretchfactor\fontdimen3\font minus
  \fontdimen4\font\relax}
\providecommand{\BIBforeignlanguage}[2]{{%
\expandafter\ifx\csname l@#1\endcsname\relax
\typeout{** WARNING: IEEEtran.bst: No hyphenation pattern has been}%
\typeout{** loaded for the language `#1'. Using the pattern for}%
\typeout{** the default language instead.}%
\else
\language=\csname l@#1\endcsname
\fi
#2}}
\providecommand{\BIBdecl}{\relax}
\BIBdecl

\bibitem{5G_Mas_MIMO_6}
M.~{Shafi}, A.~F. {Molisch}, P.~J. {Smith}, T.~{Haustein}, P.~{Zhu}, P.~{De
  Silva}, F.~{Tufvesson}, A.~{Benjebbour}, and G.~{Wunder}, ``{5G}: A tutorial
  overview of standards, trials, challenges, deployment, and practice,''
  \emph{IEEE J. Sel. Areas Commun.}, vol.~35, no.~6, pp. 1201--1221, June 2017.

\bibitem{5G_Mas_MIMO_mmWave_5}
S.~A. {Busari}, K.~M.~S. {Huq}, S.~{Mumtaz}, L.~{Dai}, and J.~{Rodriguez},
  ``Millimeter-wave massive {MIMO} communication for future wireless systems: A
  survey,'' \emph{IEEE Commun. Surveys Tuts.}, vol.~20, no.~2, pp. 836--869,
  2nd Quart. 2018.

\bibitem{5G_Book}
X.~Lin and N.~Lee, \emph{5G and Beyond}.\hskip 1em plus 0.5em minus 0.4em\relax
  Springer, 2021.

\bibitem{Report_5G_Rel17_Mt_256}
{3GPP TR 38.913}, ``5{G}: Study on scenarios and requirements for next
  generation access technologies,'' Tech. Rep. Ver. 17.0.0, May 2022.

\bibitem{6G_Docomo}
{NTT DOCOMO}, ``White paper: {5G} evolution and {6G} ({V}er. 4.0),'' pp. 1--60,
  Jan. 2022.

\bibitem{mmWave_Survey}
R.~W. {Heath}, N.~{González-Prelcic}, S.~{Rangan}, W.~{Roh}, and A.~M.
  {Sayeed}, ``An overview of signal processing techniques for millimeter wave
  {MIMO} systems,'' \emph{IEEE J. Sel. Topics Signal Process.}, vol.~10, no.~3,
  pp. 436--453, Apr. 2016.

\bibitem{6G_prospective_look}
L.~{Bariah}, L.~{Mohjazi}, S.~{Muhaidat}, P.~C. {Sofotasios}, G.~K. {Kurt},
  H.~{Yanikomeroglu}, and O.~A. {Dobre}, ``A prospective look: Key enabling
  technologies, applications and open research topics in {6G} networks,''
  \emph{IEEE Access}, vol.~8, pp. 174\,792--174\,820, 2020.

\bibitem{MassMIMO_Precoding_2021}
M.~A. Albreem, A.~H. Al~Habbash, A.~M. Abu-Hudrouss, and S.~S. Ikki, ``Overview
  of precoding techniques for massive {MIMO},'' \emph{IEEE Access}, vol.~9, pp.
  60\,764--60\,801, 2021.

\bibitem{Mass_MIMO_Hybrid_Survey}
A.~F. Molisch, V.~V. Ratnam, S.~Han, Z.~Li, S.~L.~H. Nguyen, L.~Li, and
  K.~Haneda, ``Hybrid beamforming for massive {MIMO}: A survey,'' \emph{IEEE
  Commun. Mag.}, vol.~55, no.~9, pp. 134--141, Sept. 2017.

\bibitem{Mass_MIMO_Hyb_Survey}
I.~{Ahmed}, H.~{Khammari}, A.~{Shahid}, A.~{Musa}, K.~S. {Kim}, E.~{De
  Poorter}, and I.~{Moerman}, ``A survey on hybrid beamforming techniques in
  5{G}: Architecture and system model perspectives,'' \emph{IEEE Commun.
  Surveys Tuts.}, vol.~20, no.~4, pp. 3060--3097, 4th Quart. 2018.

\bibitem{Mass_MIMO_Hybrid_Survey_2}
M.~{Rihan}, T.~{Abed Soliman}, C.~{Xu}, L.~{Huang}, and M.~I. {Dessouky},
  ``Taxonomy and performance evaluation of hybrid beamforming for {5G} and
  beyond systems,'' \emph{IEEE Access}, vol.~8, pp. 74\,605--74\,626, Mar.
  2020.

\bibitem{MassMIMO_hybrid_NO_ADMA}
H.~{Lin}, F.~{Gao}, S.~{Jin}, and G.~Y. {Li}, ``A new view of multi-user hybrid
  massive {MIMO}: Non-orthogonal angle division multiple access,'' \emph{IEEE
  J. Sel. Areas Commun.}, vol.~35, no.~10, pp. 2268--2280, Oct. 2017.

\bibitem{JSDM_LargeArray}
A.~Adhikary, J.~Nam, J.~Y. Ahn, and G.~Caire, ``Joint spatial division and
  multiplexing--the large-scale array regime,'' \emph{IEEE Trans. Inf. Theory},
  vol.~59, no.~10, pp. 6441--6463, Oct. 2013.

\bibitem{ASIL_EE_2D_OJ_COMS}
M.~Mahmood, A.~Koc, and T.~Le-Ngoc, ``Energy-efficient {MU}-massive-{MIMO}
  hybrid precoder design: Low-resolution phase shifters and digital-to-analog
  converters for {2D} antenna array structures,'' \emph{IEEE Open J. Commun.
  Soc.}, vol.~2, pp. 1842--1861, 2021.

\bibitem{ASIL_ABHP_Access}
A.~{Koc}, A.~{Masmoudi}, and T.~{Le-Ngoc}, ``{3D} angular-based hybrid
  precoding and user grouping for uniform rectangular arrays in massive
  {MU-MIMO} systems,'' \emph{IEEE Access}, vol.~8, pp. 84\,689--84\,712, May
  2020.

\bibitem{ASIL_FD_MU_NOBF}
A.~Koc and T.~Le-Ngoc, ``Intelligent non-orthogonal beamforming with large
  self-interference cancellation capability for full-duplex multiuser massive
  {MIMO} systems,'' \emph{IEEE Access}, vol.~10, pp. 51\,771--51\,791, 2022.

\bibitem{Fairness_Survey_1}
W.~Ogryczak, H.~Luss, M.~Pi{\'o}ro, D.~Nace, and A.~Tomaszewski, ``Fair
  optimization and networks: A survey,'' \emph{J. Appl. Math.}, vol. 2014,
  2014.

\bibitem{Fairness_Survey_2}
S.~Huaizhou, R.~V. Prasad, E.~Onur, and I.~Niemegeers, ``Fairness in wireless
  networks: Issues, measures and challenges,'' \emph{IEEE Commun. Surveys
  Tuts.}, vol.~16, no.~1, pp. 5--24, 2013.

\bibitem{MaxMinFairness}
M.~Sadeghi, E.~Bj{\"o}rnson, E.~G. Larsson, C.~Yuen, and T.~L. Marzetta,
  ``Max--min fair transmit precoding for multi-group multicasting in massive
  {MIMO},'' \emph{IEEE Trans. Wireless Commun.}, vol.~17, no.~2, pp.
  1358--1373, 2017.

\bibitem{MMF_NOMA_Melda_2021}
A.~Z. Yalcin, M.~K. Cetin, and M.~Yuksel, ``Max-min fair precoder design and
  power allocation for {MU-MIMO NOMA},'' \emph{IEEE Trans. Veh. Technol.},
  vol.~70, no.~6, pp. 6217--6221, 2021.

\bibitem{MaxMin_HP_Maritime}
C.~Liu, W.~Feng, T.~Wei, and N.~Ge, ``Fairness-oriented hybrid precoding for
  massive {MIMO} maritime downlink systems with large-scale {CSIT},''
  \emph{China Commun.}, vol.~15, no.~1, pp. 52--61, 2018.

\bibitem{proportional_Fairness}
Y.~Lin, Y.~Wang, C.~Li, Y.~Huang, and L.~Yang, ``Joint design of user
  association and power allocation with proportional fairness in massive {MIMO
  HetNets},'' \emph{IEEE Access}, vol.~5, pp. 6560--6569, 2017.

\bibitem{HBF_fairness_proportional}
I.~Ahmed, H.~Khammari, and A.~Shahid, ``Resource allocation for transmit hybrid
  beamforming in decoupled millimeter wave multiuser-{MIMO} downlink,''
  \emph{IEEE Access}, vol.~5, pp. 170--182, 2017.

\bibitem{SumRateMax_NOMA_LEO}
Z.~Gao, A.~Liu, C.~Han, and X.~Liang, ``Sum rate maximization of massive {MIMO
  NOMA in LEO} satellite communication system,'' \emph{IEEE Wireless Commun.
  Lett.}, vol.~10, no.~8, pp. 1667--1671, 2021.

\bibitem{ASIL_PSO_PA_WCNC}
A.~{Koc} and T.~{Le-Ngoc}, ``Swarm intelligence based power allocation in
  hybrid massive {MIMO} systems,'' in \emph{2021 IEEE Wireless Commun. and
  Netw. Conf. (WCNC)}, Mar. 2021, pp. 1--7.

\bibitem{ASIL_DL_PA_ICC}
A.~{Koc}, M.~{Wang}, and T.~{Le-Ngoc}, ``Deep learning based multi-user power
  allocation and hybrid precoding in massive {MIMO} systems,'' in \emph{2022
  IEEE Int. Conf. Commun. (ICC 2022)}, May 2022, pp. 1--6.

\bibitem{ASIL_GA_RA_OFDM_WCNC}
A.~{Koc}, F.~{Bishe}, and T.~{Le-Ngoc}, ``Energy-efficient throughput
  maximization in {mmWave MU-Massive-MIMO-OFDM}: Genetic algorithm based
  resource allocation,'' in \emph{2022 IEEE Wireless Commun. and Netw. Conf.
  (WCNC)}, Apr. 2022, pp. 1--6.

\bibitem{alphaFAIR_1_NOMA}
P.~Xu and K.~Cumanan, ``Optimal power allocation scheme for non-orthogonal
  multiple access with $\alpha$-fairness,'' \emph{IEEE J. Sel. Areas Commun.},
  vol.~35, no.~10, pp. 2357--2369, 2017.

\bibitem{PSO_AI}
W.~{Tong}, A.~{Hussain}, W.~X. {Bo}, and S.~{Maharjan}, ``Artificial
  intelligence for vehicle-to-everything: A survey,'' \emph{IEEE Access},
  vol.~7, pp. 10\,823--10\,843, 2019.

\bibitem{PSO_book}
X.-S. Yang, \emph{Nature-inspired optimization algorithms}.\hskip 1em plus
  0.5em minus 0.4em\relax Elsevier, 2014.

\bibitem{GWO_first_paper}
S.~Mirjalili, S.~M. Mirjalili, and A.~Lewis, ``Grey wolf optimizer,''
  \emph{Adv. Eng. Softw.}, vol.~69, pp. 46--61, 2014.

\bibitem{ACO_cont}
K.~Socha and M.~Dorigo, ``Ant colony optimization for continuous domains,''
  \emph{European J. Oper. Res.}, vol. 185, no.~3, pp. 1155--1173, 2008.

\bibitem{NFL_first_TEVC}
D.~H. Wolpert and W.~G. Macready, ``No free lunch theorems for optimization,''
  \emph{IEEE Transa. Evol. Comput.}, vol.~1, no.~1, pp. 67--82, 1997.

\bibitem{UAV_swarm_intelligence}
D.~Pliatsios, S.~K. Goudos, T.~Lagkas, V.~Argyriou, A.-A.~A. Boulogeorgos, and
  P.~Sarigiannidis, ``Drone-base-station for next-generation
  internet-of-things: A comparison of swarm intelligence approaches,''
  \emph{IEEE Open J. Antennas Propag.}, vol.~3, pp. 32--47, 2022.

\bibitem{ChannelModels}
X.~{Cheng}, B.~{Yu}, L.~{Yang}, J.~{Zhang}, G.~{Liu}, Y.~{Wu}, and L.~{Wan},
  ``Communicating in the real world: {3D MIMO},'' \emph{IEEE Wireless Commun.},
  vol.~21, no.~4, pp. 136--144, Aug. 2014.

\bibitem{COST_2100_MIMO}
L.~Liu, C.~Oestges, J.~Poutanen, K.~Haneda, P.~Vainikainen, F.~Quitin,
  F.~Tufvesson, and P.~De~Doncker, ``The {COST 2100 MIMO} channel model,''
  \emph{IEEE Wireless Commun.}, vol.~19, no.~6, pp. 92--99, 2012.

\bibitem{GBSM_3GPP}
J.~Yang, B.~Ai, K.~Guan, D.~He, X.~Lin, B.~Hui, J.~Kim, and A.~Hrovat, ``A
  geometry-based stochastic channel model for the millimeter-wave band in a
  {3GPP} high-speed train scenario,'' \emph{IEEE Trans. Veh. Technol.},
  vol.~67, no.~5, pp. 3853--3865, 2018.

\bibitem{Emil_PA_NonConvex}
E.~Bj{\"o}rnson, M.~Bengtsson, and B.~Ottersten, ``Optimal multiuser transmit
  beamforming: A difficult problem with a simple solution structure [lecture
  notes],'' \emph{IEEE Signal Process. Mag.}, vol.~31, no.~4, pp. 142--148,
  2014.

\bibitem{PA_nonConvex}
Y.-F. Liu, Y.-H. Dai, and Z.-Q. Luo, ``Coordinated beamforming for {MISO}
  interference channel: Complexity analysis and efficient algorithms,''
  \emph{IEEE Trans. Signal Process.}, vol.~59, no.~3, pp. 1142--1157, 2011.

\bibitem{PA_nonConvex_2}
C.~Pan, H.~Ren, M.~Elkashlan, A.~Nallanathan, and L.~Hanzo, ``The non-coherent
  ultra-dense {C-RAN} is capable of outperforming its coherent counterpart at a
  limited fronthaul capacity,'' \emph{IEEE J. Sel. Areas Commun.}, vol.~36,
  no.~11, pp. 2549--2560, 2018.

\bibitem{AoD_Est_2_Decades}
H.~{Krim} and M.~{Viberg}, ``Two decades of array signal processing research,''
  \emph{IEEE Signal Process. Mag.}, vol.~13, no.~4, pp. 67--94, July 1996.

\bibitem{ASIL_Xiaoyi_DL_CE}
X.~Zhu, A.~Koc, R.~Morawski, and T.~Le-Ngoc, ``A deep learning and geospatial
  data based channel estimation technique for hybrid massive {MIMO} systems,''
  \emph{IEEE Access}, vol.~9, pp. 145\,115--145\,132, 2021.

\bibitem{yu2007transmitter}
W.~Yu and T.~Lan, ``Transmitter optimization for the multi-antenna downlink
  with per-antenna power constraints,'' \emph{IEEE Trans. Signal Process.},
  vol.~55, no.~6, pp. 2646--2660, 2007.

\bibitem{Report_5G_Macro_PL_Rel_17}
{3GPP TR 36.931}, ``{LTE}; evolved universal terrestrial radio access
  ({E-UTRA}); radio frequency ({RF}) requirements for {LTE} pico node {B},''
  Tech. Rep. Ver. 17.0.0, Apr. 2022.

\bibitem{Report_5G_UMi_UMa_Rel17}
{3GPP TR 38.901}, ``5{G}: Study on channel model for frequencies from 0.5 to
  100 {GHz},'' Tech. Rep. Ver. 17.0.0, Apr. 2022.

\bibitem{ANALOG_BF_Heath}
X.~{Gao}, L.~{Dai}, S.~{Han}, C.~{I}, and R.~W. {Heath}, ``Energy-efficient
  hybrid analog and digital precoding for mmwave {MIMO} systems with large
  antenna arrays,'' \emph{IEEE J. Sel. Areas Commun.}, vol.~34, no.~4, pp.
  998--1009, Apr. 2016.

\bibitem{strongDuality_Lagrangian}
S.~Wolf and S.~M. G{\"u}nther, ``An introduction to duality in convex
  optimization,'' \emph{Netw.}, vol. 153, 2011.

\end{thebibliography}
	
	
%


\begin{IEEEbiography}[{\includegraphics[width=1in,height=1.25in,clip,keepaspectratio]{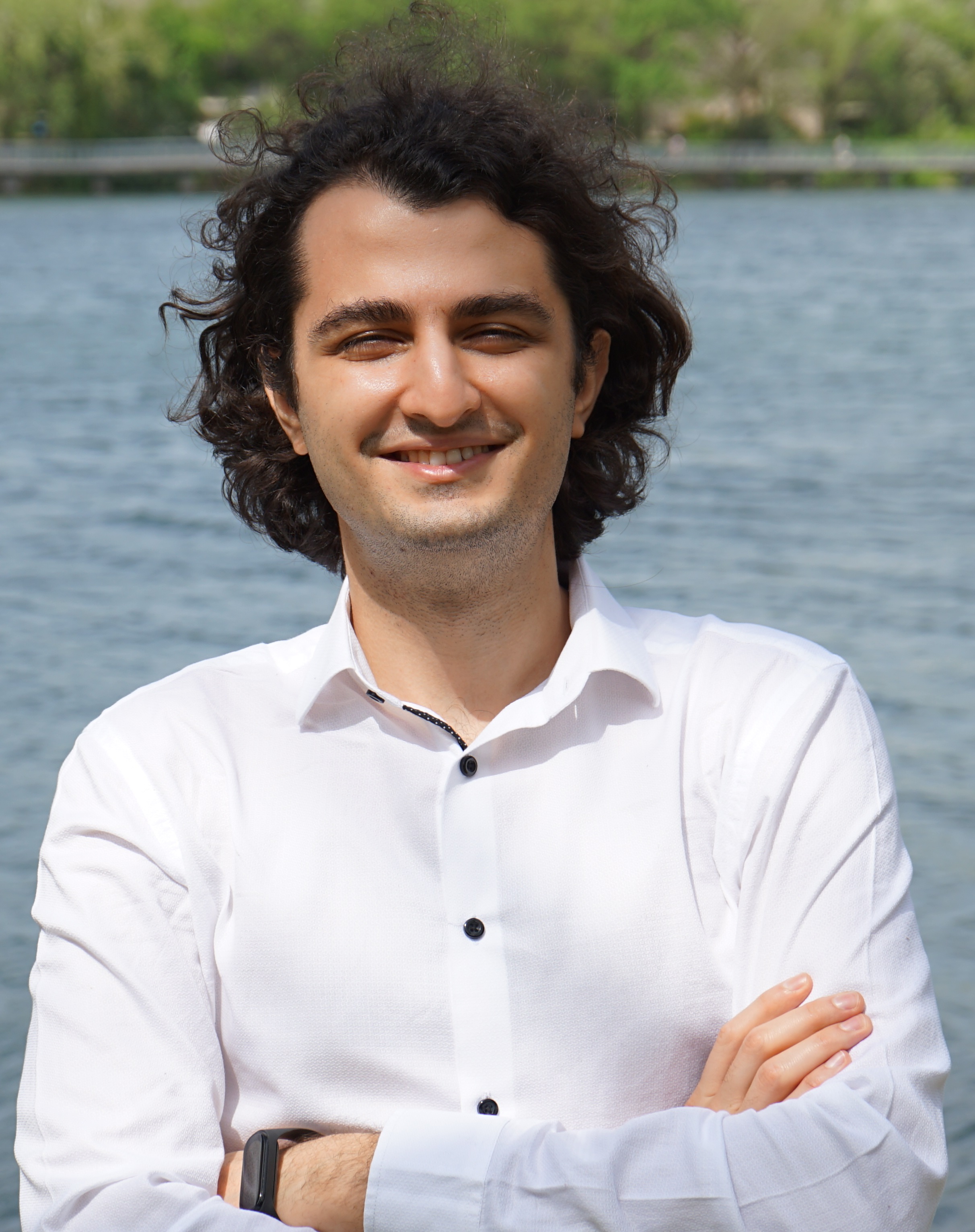}}]{Asil Koc}
	(Graduate Student Member, IEEE) received the B.Sc. degree (Hons.) in electronics and communication engineering, the M.Sc. degree (Hons.) in telecommunication engineering from Istanbul Technical University, Istanbul, Turkey, in 2015 and 2017, respectively. Currently, he is a Ph.D. candidate in electrical engineering at
	McGill University, Montreal, Canada. 
	
	From 2015 to 2017, he was a Research and Teaching Assistant with the Electronics and Communication Engineering Department, Istanbul Technical University. 
		Since 2017 and 2021, he has been a Teaching Assistant and Lecturer, respectively, with the Electrical and Computer Engineering Department, McGill University.
	His research interests include, but not limited to AI/ML based wireless communications, massive MIMO, full-duplex, millimeter-wave/terahertz, beamforming, index modulation, wireless power transfer, and cooperative networks. He was a recipient of 
	Erasmus Scholarship by European Union, 
	McGill Engineering Doctoral Award, 
	IEEE ComSoc Student Travel Grant,
	Graduate Research Enhancement and Travel Award by McGill University, and
	STARaCom Collaborative Grant by the FRQNT.
\end{IEEEbiography}

%

\begin{IEEEbiography}[{\includegraphics[width=1in,height=1.25in,clip,keepaspectratio]{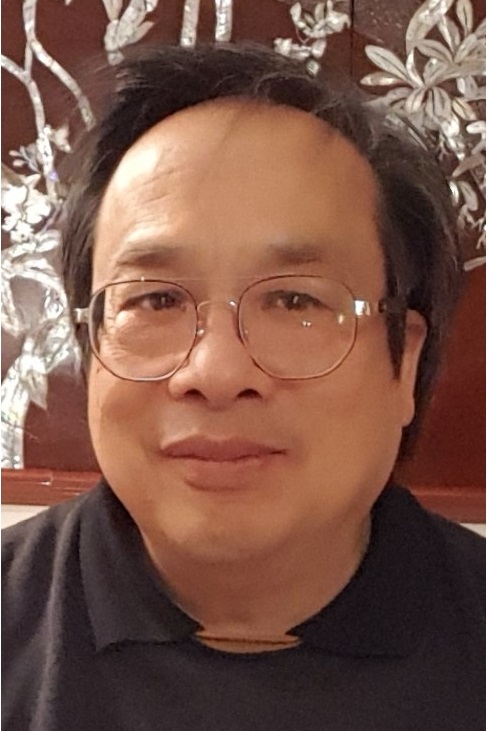}}]{Tho Le-Ngoc}
	(Life Fellow, IEEE) received the B.Eng. degree in electrical engineering, in 1976, the M.Eng. degree in microprocessor applications, in 1978, from McGill University, Montreal, and the Ph.D. degree in digital communications, in 1983, from the University of Ottawa, Canada. 
	
	From 1977 to 1982, he was with Spar Aerospace Ltd., Sainte-Anne-de-Bellevue, QC, Canada, involved in the development and design of satellite communications systems. From 1982 to 1985, he was with SRTelecom Inc., Saint-Laurent, QC, Canada, where he developed the new point-to-multipoint DA-TDMA/TDM Subscriber Radio System SR500. From 1985 to 2000, he was a Professor with the Department of Electrical and Computer Engineering, Concordia University, Montreal. Since 2000, he has been with the Department of Electrical and Computer Engineering, McGill University. His research interest includes broadband digital communications. He is  {a Distinguished James McGill Professor, and} a Fellow of the Engineering Institute of Canada, the Canadian Academy of Engineering, and the Royal Society of Canada. He was a recipient of the 2004 Canadian Award in Telecommunications Research and the IEEE Canada Fessenden Award, in 2005. 
\end{IEEEbiography}

	
\end{document}